\documentclass{article}\usepackage[]{graphicx}\usepackage[]{color}
\makeatletter
\def\maxwidth{ %
  \ifdim\Gin@nat@width>\linewidth
    \linewidth
  \else
    \Gin@nat@width
  \fi
}
\makeatother

\definecolor{fgcolor}{rgb}{0.345, 0.345, 0.345}

\usepackage{framed}
\makeatletter
 {\par\unskip\endMakeFramed%
 \at@end@of@kframe}
\makeatother

\definecolor{shadecolor}{rgb}{.97, .97, .97}
\definecolor{messagecolor}{rgb}{0, 0, 0}
\definecolor{warningcolor}{rgb}{1, 0, 1}
\definecolor{errorcolor}{rgb}{1, 0, 0}
\newenvironment{knitrout}{}{} 

\usepackage{alltt}

\usepackage{arxiv}

\RequirePackage{amsthm,amsmath,dsfont,amssymb,array,arydshln, booktabs}
\usepackage{xcolor, colortbl}
\usepackage{hyperref}

\title{Combining heterogeneous subgroups with \\ graph-structured variable selection priors \\ for Cox regression}

\author{
  Katrin~Madjar \\
  Department of Statistics\\
	TU Dortmund University\\
	44221 Dortmund, Germany \\
  \texttt{madjar@statistik.tu-dortmund.de} \\
   \And
  Manuela~Zucknick \\
  Oslo Centre for Biostatistics and Epidemiology\\
  Department of Biostatistics\\ 
  University of Oslo\\ 
  0317 Oslo, Norway \\
  \texttt{manuela.zucknick@medisin.uio.no} \\
   \And
 Katja~Ickstadt \\
 Department of Statistics\\
	TU Dortmund University\\
	44221 Dortmund, Germany \\
  \texttt{ickstadt@statistik.tu-dortmund.de} \\
   \And
 J\"org~Rahnenf\"uhrer \\
 Department of Statistics\\
	TU Dortmund University\\
	44221 Dortmund, Germany \\
  \texttt{rahnenfuehrer@statistik.tu-dortmund.de} \\
}

\makeatletter
\g@addto@macro\appendix{\setcounter{figure}{0}}
\g@addto@macro\appendix{\setcounter{table}{0}}
\makeatother
\IfFileExists{upquote.sty}{\usepackage{upquote}}{}
\begin{document}
\maketitle

\begin{abstract}

Important objectives in cancer research are the prediction of a patient's risk based on molecular measurements such as gene expression data and the identification of new prognostic biomarkers (e.g. genes). In clinical practice, this is often challenging because patient cohorts are typically small and can be heterogeneous. 
In classical subgroup analysis, a separate prediction model is fitted using only the data of one specific cohort. However, this can lead to a loss of power when the sample size is small. Simple pooling of all cohorts, on the other hand, can lead to biased results, especially when the cohorts are heterogeneous.
For this situation, 
we propose a new Bayesian approach suitable for continuous molecular measurements and survival outcome that identifies the important predictors and provides a separate risk prediction model for each cohort. 
It allows sharing information between cohorts to increase power by assuming a graph linking predictors within and across different cohorts. 
The graph helps to identify pathways of functionally related genes and genes that are simultaneously prognostic in different cohorts.
Results demonstrate that our proposed approach is superior to the standard approaches in terms of prediction performance and increased power in variable selection when the sample size is small.

\end{abstract}


\section{Introduction}

In clinical research, molecular measurements such as gene expression data play an important role in the diagnosis and prediction of a disease outcome, such as time-to-event endpoint.
In general, the number of molecular predictors is larger than the sample size (``$p>n$ problem'') and typically only a small number of genes is associated with the outcome while the rest is noise. 
Thus, important objectives in statistical modeling are good prediction performance and variable selection to obtain a subset of prognostic predictors.

In the Bayesian framework, different types of variable selection priors have been proposed also with application to the Bayesian Cox model. 
One common choice is the use of shrinkage priors such as the Bayesian lasso as an analog to the frequentist penalized likelihood approach \cite{park_bayesian_2008, lee_bayesian_2011, zucknick_nonidentical_2015}.
A popular alternative are ``spike-and-slab'' priors that use latent indicators for variable selection and a mixture distribution for the regression coefficients 
\cite{george_variable_1993, treppmann_integration_2017}. 
In general, the regression coefficients are modeled independently.
However, with applications to molecular data, it can be reasonable to consider structural information between covariates, since the effect on a clinical outcome is typically not caused by single genes acting in isolation, but rather by changes in a regulatory or functional pathway of interacting genes.
Several authors have dealt with this problem by using a Markov random field (MRF) prior to incorporate structural information on the relationships among the covariates into variable selection \cite{li_bayesian_2010, stingo_variable_2011, stingo_incorporating_2011, peterson_joint_2016}.
Alternatively, \cite{Chakraborty_agraph_2019} propose a Graph Laplacian prior for modeling the dependence structure between the regression coefficients through their precision matrix.

When the data are heterogeneous and consists of known subpopulations with possibly different dependence structures, 
estimating one joint graphical model would hide the underlying heterogeneity while estimating separate models for each subpopulation would neglect common structure.
For this situation, \cite{danaher_joint_2014} use an extension of the frequentist graphical lasso with either a group or fused lasso type penalty for joint structure learning. 
\cite{saegusa_joint_2016} propose a weighted Laplacian shrinkage penalty where the weights represent the degree of sim\-i\-lar\-i\-ty between subpopulations. 
Bayesian approaches for sharing common structure in the joint inference of multiple graphical models have also been developed \cite{yajima_differential_2012, mitra_bayesian_2016, peterson_bayesian_2015}.
\cite{peterson_bayesian_2015} use an MRF prior for the graph structures with pairwise similarities between different graphs. 
However, all these methods have in common that they focus on structure learning only and do not take into account the relationship between (structured) covariates and a clinical outcome as in the context of regression modeling. 

We consider the situation that molecular measurements and a survival outcome are available for different, possibly heterogeneous patient subgroups or cohorts such as in a multicenter study. 
In classical subgroup analysis, only the data of the subgroup of interest is used to build a risk prediction model for this specific subgroup.
This may lead to a loss of power or unstable results with high variance especially in small subgroups.
Thus, it is tempting to simply pool all data to increase the sample size.
This approach, however, can result in biased estimates when the subgroups are heterogeneous regarding their effects and subgroup-specific effects may get lost.
We aim at sharing information between subgroups to increase power when this is supported by the data.
Our approach provides a separate risk prediction model for each subgroup that allows the identification of common as well as subgroup-specific effects and has improved prediction accuracy and variable selection power compared to the two standard approaches.

Some frequentist approaches tackle this problem by suggesting a penalized Cox regression model with a weighted version of the partial likelihood that includes patients of all subgroups but assigns them (individual) weights. 
\cite{weyer_weighting_2015} propose the use of fixed weights. 
This idea is extended by \cite{Richter_MBO_2019} using model-based optimization for tuning of the weights to obtain the best combination of fixed weights regarding prediction accuracy.
\cite{madjar_weighted_2020} estimate individual weights from the data such that they represent the probability of belonging to a specific subgroup. 

In this paper, we use a Bayesian approach and borrow information across subgroups through graph-structured selection priors instead of weights in the likelihood.
We propose an extension of the Bayesian Cox model with ``spike-and-slab'' prior for variable selection by \cite{treppmann_integration_2017} in the sense that we incorporate graph information between covariates into variable selection via an MRF prior instead of modeling the regression coefficients independently.
The graph is not known a priori and inferred simultaneously with the important predictors. 
Its structure can be partitioned into subgraphs linking covariates within or across different subgroups.
Thus, representing conditional dependencies between genes (i.e. pathways) and similarities between subgroups by genes being simultaneously prognostic in different subgroups.

The paper is structured as follows: the statistical methods are described in section~\ref{sec:statmeth}, first in the general form and then adapted to our situation.
Section~\ref{sec:simstudy} covers the simulation setup along with the simulation results.
A case study with Glioblastoma protein expression data is provided in section~\ref{sec:gbd}.
The paper concludes with a summary and discussion of the main findings in section~\ref{sec:discuss}. 

\section{Statistical Methods}
\label{sec:statmeth}

First, the general methods are described that are required for our proposed Bayesian model introduced in section \ref{sec:propmodel}.

\subsection{The Bayesian Cox proportional hazards model}
\label{sec:coxmodel}

Assume the observed data of patient $m$ consist of the tuple $(\tilde{t}_m, \delta_m)$ and the covariate vector ${\boldsymbol{x}_m=(x_{m1},\ldots,x_{mp})' \in \mathds{R}^{p}}$, 
$m=1,\ldots,n$. 
$\boldsymbol{x}\in \mathds{R}^{n\times p}$ is the matrix of (genomic) covariates.
$\tilde{t}_m=\min(T_m,C_m)$ denotes the observed time of patient $m$, with $T_m$ the event time and $C_m$ the censoring time.
$\delta_m = \mathds{1}(T_m \leq C_m)$ indicates whether a patient experienced an event ($\delta_m=1$) or was right-censored ($\delta_m=0$).

The Cox proportional hazards model \cite{cox_regression_1972} models the hazard rate $h(t|\boldsymbol{x}_m)$ of an individual $m$ at time $t$. 
It consists of two terms, the non-parametric baseline hazard rate $h_0(t)$ and a parametric form of the covariate effect: 
\begin{equation*}
h(t|\boldsymbol{x}_m)= h_0(t) \cdot \exp(\boldsymbol{\beta}' \boldsymbol{x}_m ) = h_0(t) \cdot \exp \left(\sum_{i=1}^p \beta_i x_{mi} \right) ,
\end{equation*}
where $\boldsymbol{\beta}=(\beta_1,...,\beta_p)'$ is the unknown parameter vector that represents the strength of influence of the covariates on the hazard rate. 

Under the Cox model, the joint survival probability of $n$ patients given $\boldsymbol{x}$ is 
\[
P(\tilde{\boldsymbol{T}}>\tilde{\boldsymbol{t}}|\boldsymbol{x}, \boldsymbol{\beta}, H_0) = \exp\Big( - \sum_{m=1}^n \exp(\boldsymbol{\beta}'\boldsymbol{x}_m) H_0(\tilde{t}_m) \Big) .
\]
One of the most popular choices for the cumulative baseline hazard function $H_0(t)$ is a gamma process prior 
\[
H_0 \sim \mathcal{GP}(a_0H^\ast, a_0) ,
\]
where $H^\ast(t)$ is an increasing function with $H^\ast(0)=0$. $H^\ast$ can be considered as an initial guess of $H_0$ and $a_0>0$ describes the weight that is given to $H^\ast(t)$ {\cite{lee_bayesian_2011}}. 
\cite{lee_bayesian_2011} propose a Weibull distribution $H^\ast(t) = \eta t^{\kappa}$ with fixed hyperparameters $\eta$ and $\kappa$.
Following \cite{zucknick_nonidentical_2015}, we obtain estimates of $\eta$ and $\kappa$ from the training data by fitting a parametric Weibull model without covariates to the survival data. 
We choose $a_0=2$ in accordance with the authors. 

In practice the presence of ties is very common, leading to the grouped data likelihood described in \cite[chapter 3.2.2]{ibrahim_bayesian_2005}. 
A finite partition of the time axis is constructed with ${0=c_0<c_1<...<c_J}$ and $c_J>\tilde{t}_m$ for all $m=1,...,n$. The observed time $\tilde{t}_m$ of patient $m$ falls in one of the $J$ disjoint intervals $I_g=(c_{g-1},c_g], g=1,...,J$. Assume the observed data $\mathfrak{D}=\{(\boldsymbol{x}, \mathcal{R}_g, \mathcal{D}_g): g=1,...,J\}$ are grouped within $I_g$, where $\mathcal{R}_g$ and $\mathcal{D}_g$ are the risk and failure sets corresponding to interval $g$.
Let $h_g = H_0(c_g) - H_0(c_{g-1})$ be the increment in the cumulative baseline hazard in interval $I_g$, $g=1,...,J$. From the gamma process prior of $H_0$ follows that the $h_g$'s have independent gamma distributions
\[
h_g \sim \mathcal{G}(\alpha_{0,g}-\alpha_{0,g-1}, a_0) \, , \quad \text{with} \quad \alpha_{0,g}=a_0H^\ast(c_g) \, .
\]
The conditional probability 
that the observed time of patient $m$ falls in interval $I_g$ is given by
\begin{align*}
P(\tilde{T}_m \in I_g|\boldsymbol{h}) &= \exp \Big( -\exp(\boldsymbol{\beta}'\boldsymbol{x}_m) \sum_{j=1}^{g-1} h_j \Big) \cdot \Big[ 1 - \exp \big( -h_g \exp(\boldsymbol{\beta}'\boldsymbol{x}_m) \big) \Big],
\end{align*}
with $\boldsymbol{h} = (h_1,...,h_J)'$.
The resulting grouped data likelihood is defined as
\[
L(\mathfrak{D}|\boldsymbol{\beta}, \boldsymbol{h}) \propto 
\prod_{g=1}^J \left[ \exp\Big( -h_g \! \! \sum_{k \in \mathcal{R}_g-\mathcal{D}_g} \! \! \exp(\boldsymbol{\beta}'\boldsymbol{x}_k) \Big) \prod_{l \in \mathcal{D}_g} \Big[ 1-\exp\big( -h_g \exp(\boldsymbol{\beta}' \boldsymbol{x}_l) \big) \Big] \right]
\]
\cite[chapter 3.2.2]{ibrahim_bayesian_2005}.

\subsection{Stochastic search variable selection}  
\label{sec:ssvs}


The stochastic search variable selection (SSVS) procedure by \cite{george_variable_1993} uses latent indicators for variable selection and models the regression coefficients as a mixture of two normal distributions with different variances
\[
\beta_i|\gamma_i \sim (1-\gamma_i) \cdot \mathcal{N}(0, \tau_i^2) + \gamma_i \cdot \mathcal{N}(0, c_i^2 \tau_i^2) \, , \quad i=1,...,p \, .
\]
This prior allows the $\beta_i$'s to shrink towards zero. Due to the shape of the two-component mixture distribution, it is called \textit{spike-and-slab prior}. 
The latent variable $\gamma_i$ indicates the inclusion ($\gamma_i=1$) or exclusion ($\gamma_i=0$) of the $i$-th variable and specifies the variance of the normal distribution.
$\tau_i~(>0)$ is set small so that $\beta_i$ is likely to be close to zero if $\gamma_i=0$. 
$c_i~(>1)$ is chosen sufficiently large to inflate the coefficients of selected variables and to make their posterior mean values likely to be non-zero. 
In general, the variances of the regression coefficients are assumed to be constant: $\tau_i \equiv \tau$ and $c_i \equiv c$ for all $i=1,...,p$.
%

The standard prior for $\boldsymbol{\gamma}=(\gamma_1,...,\gamma_p)'$ is a product of independent Bernoulli distributions  
\[
p(\boldsymbol{\gamma}) = \prod_{i=1}^p \pi^{\gamma_i} \cdot (1-\pi)^{1-\gamma_i}  ,
\]
with prior inclusion probability $\pi=P(\gamma_i=1)$.
Typically, these prior inclusion probabilities are chosen to be the same for all variables and often with $\pi$ set to a fixed value. 

\subsection{Graphical models}
\label{sec:bayesnetzw}

A graphical model is a statistical model that is associated with a graph summarizing the dependence structure in the data. 
The nodes of a graph represent the random variables of interest and the edges of a graph describe conditional dependencies among the variables. Structure learning implies the estimation of an unknown graph. 
Recent applications are mainly driven by biological problems that involve the reconstruction of gene regulatory networks and the identification of pathways of functionally related genes from their expression levels.
A graph is called \textit{undirected}, when its edges are unordered pairs of nodes instead of ordered pairs with edges pointing from one node to the other (\textit{directed} graph).
When the variables are continuous measurements and assumed to be multivariate normal a common choice are Gaussian models \cite{drton_structure_2017}.

We assume that the vector of random variables $\boldsymbol{X}_m = (X_{m1},...,X_{mp})'$ for patient $m$, $m=1,...,n$ 
follows a multivariate normal distribution with mean vector $\boldsymbol{0}$ and covariance matrix $\boldsymbol{\Sigma}$.
The inverse of the covariance matrix is referred to as precision 
matrix $\boldsymbol{\Sigma}^{-1}=\boldsymbol{\Omega}=(\omega_{ij})_{i,j=1,...,p}$, with $\boldsymbol{\Omega}$ symmetric and positive definite.
Let $\boldsymbol{X} \in \mathds{R}^{n \times p}$ be the data matrix consisting of $n$ independent patients and $\boldsymbol{S}=\frac{1}{n}\boldsymbol{X}'\boldsymbol{X}$ the sample covariance matrix.

In graphical models, a graph $\widetilde{G}$ is used to represent conditional dependence relationships among random variables $\boldsymbol{X}$.
Let $\widetilde{G}=(V,E)$ be an undirected graph
, where $V=\{ 1,...,p\}$ is a set of nodes (e.g. genes) and $E \subset V \times V$ is a set of edges (e.g. relations between genes) with edge ${(i,j)\in E \Leftrightarrow (j,i)\in E}$.
$\widetilde{G}$ can be indexed by a set of $p(p-1)/2$ binary variables ${\boldsymbol{G}=(g_{ij})_{i<j} \in \{0,1\}^{p \times p}}$ with $g_{ij}=1$ or 0 when edge $(i,j)$ belongs to $E$ or not. The symmetric matrix $\boldsymbol{G}$ is termed adjacency matrix representation of the graph.
The graph structure implies constraints on the precision matrix $\boldsymbol{\Omega}$ such that $\, g_{ij}=0 \, \Leftrightarrow  \, (i,j)\notin E  \, \Leftrightarrow \, \omega_{ij}=0$, 
meaning that variables $i$ and $j$ are conditionally independent given all remaining variables
\cite{drton_structure_2017, wang_scaling_2015}. 

We use the approach for structure learning by \cite{wang_scaling_2015} that is based on continuous spike-and-slab priors for the elements of the precision matrix and latent indicators for the graph structure.
The approach induces sparsity and is efficient due to a block Gibbs sampler 
and no approximation of the normalizing constant.
The corresponding hierarchical model is defined as
\[
p(\boldsymbol{\Omega}|\boldsymbol{G},\theta) = C(\boldsymbol{G}, \nu_0, \nu_1,\lambda)^{-1} \prod_{i<j} \mathcal{N}(\omega_{ij}|0,\nu_{g_{ij}}^2) \prod_{i} \text{Exp}(\omega_{ii}|\frac{\lambda}{2}) \mathds{1}_{ \{\boldsymbol{\Omega} \in \mathcal{M}^{+}\} }
\]
\[
p(\boldsymbol{G}|\theta) = C(\theta)^{-1} C(\boldsymbol{G}, \nu_0, \nu_1,\lambda) \prod_{i<j} \big( \pi^{g_{ij}} (1-\pi)^{1-g_{ij}} \big) ,
\]
where $\theta=\{\nu_0, \nu_1,\lambda, \pi\}$ is the set of all parameters with $\nu_0 > 0$ small, $\nu_1 > 0$ large, $\lambda>0$ and $\pi \in (0,1)$. 
$\mathds{1}_{\{\Omega_s \in \mathcal{M}^{+} \}}$ restricts the prior to the space of symmetric-positive definite matrices.
A small value for $\nu_0$ ($g_{ij}=0$) means that $\omega_{ij}$ is small enough to bet set to zero. A large value for $\nu_1$ ($g_{ij}=1$) allows $\omega_{ij}$ to be substantially different from zero.
The binary latent variables ${\boldsymbol{G}=(g_{ij})_{i<j} \in \{0,1\}^{p(p-1)/2}}$ serve as edge inclusion indicators. 
\cite{wang_scaling_2015} proposes the following fixed hyperparameters $\pi=\frac{2}{p-1}$, 
$\nu_0 \geq 0.01$, $\nu_1 \leq 10$ and $\lambda=1$ resulting in good convergence. 


\subsection{The proposed Bayesian subgroup model}  
\label{sec:propmodel}

We assume the entire data consists of $S$ predefined subgroups of patients, where for each patient the subgroup membership is known.

\subsubsection{Likelihood} 

Let $\boldsymbol{X}_s \in \mathds{R}^{n_s \times p}$ be the gene expression (covariate) matrix for subgroup $s$, $s=1,...,S$, consisting of $n_s$ independent and identically distributed observations. 
For patient $m$ in subgroup $s$ the vector of random variables $\boldsymbol{X}_{s,m} = (X_{s,m1},...,X_{s,mp})'$ is assumed to follow a multivariate normal distribution with mean vector $\boldsymbol{0}$ and unknown precision matrix $\boldsymbol{\Omega}_{ss}=\boldsymbol{\Sigma}_s^{-1}$, $m=1,...,n_s$.  

We consider the outcome $\boldsymbol{Y}_s=(Y_{s,1},...,Y_{s,n_s})'$ with $Y_{s,m}=(\tilde{T}_{s,m}, \delta_{s,m})$ as well as the predictors $\boldsymbol{X}_s$, to be random variables. 
Thus, the likelihood for subgroup $s$ is the joint distribution ${p(\boldsymbol{Y}_s, \boldsymbol{X}_s)=p(\boldsymbol{Y}_s|\boldsymbol{X}_s) \cdot p(\boldsymbol{X}_s)}$.
The conditional distribution $p(\boldsymbol{Y}_s|\boldsymbol{X}_s)$ corresponds to the grouped data likelihood of the Bayesian Cox proportional hazards model in section \ref{sec:coxmodel} \cite{lee_bayesian_2011} for subgroup $s$
\[
L(\mathfrak{D}_s|\boldsymbol{\beta}_s, \boldsymbol{h}_s) \propto 
\prod_{g=1}^{J_s} \left[ \exp\Big( -h_{s,g} \mspace{-20mu} \sum_{k \in \mathcal{R}_{s,g}-\mathcal{D}_{s,g}} \mspace{-20mu} \exp(\boldsymbol{\beta}_s'\boldsymbol{x}_{s,k}) \Big) \prod_{l \in \mathcal{D}_{s,g}} \Big[ 1-\exp\big( -h_{s,g} \exp(\boldsymbol{\beta}_s' \boldsymbol{x}_{s,l} ) \big) \Big] \right] , 
\]
where $\mathfrak{D}_s=\{(\boldsymbol{x}_s, \mathcal{R}_{s,g}, \mathcal{D}_{s,g}): g=1,...,J_s\}$ are the observed data in subgroup $s$, with $\mathcal{R}_g$ the risk and $\mathcal{D}_g$ the failure sets corresponding to interval ${I_{s,g}=(c_{s,g-1},c_{s,g}]}$, ${g=1,...,J_s}$. 
The increment in the cumulative baseline hazard for subgroup $s$ in interval $I_{s,g}$ is termed ${h_{s,g} = H_0(c_{s,g}) - H_0(c_{s,g-1})}$. 
$\boldsymbol{\beta}_s$ is the $p$-dimensional vector of regression coefficients for subgroup $s$.

The marginal distribution of $\boldsymbol{X}_s$ is multivariate normal with $\boldsymbol{S}_s=\boldsymbol{X}_s' \boldsymbol{X}_s$
\[
p(\boldsymbol{X}_s|\boldsymbol{\Omega}_{ss}) 
\propto \prod_{m=1}^{n_s} |\boldsymbol{\Omega}_{ss}|^{1/2} \exp\big( -\frac{1}{2} \boldsymbol{X}_{s,m}' \boldsymbol{\Omega}_{ss} \boldsymbol{X}_{s,m} \big) 
= |\boldsymbol{\Omega}_{ss}|^{n_s/2} \exp\big( -\frac{1}{2} \underbrace{\sum_{m=1}^{n_s} \boldsymbol{X}_{s,m}' \boldsymbol{\Omega}_{ss} \boldsymbol{X}_{s,m}}_{=\text{tr}(\boldsymbol{S}_s \boldsymbol{\Omega}_{ss})} \big) . 
\]

The joint likelihood across all subgroups is the product of the subgroup likelihoods
\[
\prod_{s=1}^S L(\mathfrak{D}_s|\boldsymbol{\beta}_s, \boldsymbol{h}_s) \cdot p(\boldsymbol{X}_s|\boldsymbol{\Omega}_{ss}) .
\]

\subsubsection{Prior specifications}

\subsubsection*{ Prior on the parameters $\boldsymbol{h}_s$ and $\boldsymbol{\beta}_s$ of the Cox model}

The prior for the increment in the cumulative baseline hazard in subgroup $s$ follows independent gamma distributions
\[
h_{s,g} \sim \mathcal{G}(a_0 (H^\ast(c_{s,g})-H^\ast(c_{s,g-1})), a_0)   , 
\]
with a Weibull distribution $H^\ast(c_{s,g}) = \eta_s c_{s,g}^{\kappa_s}$, $g=1,...,J_s$, $s=1,...,S$ \cite{lee_bayesian_2011}. 
We choose the hyperparameters $a_0$, $\eta_s$ and $\kappa_s$ to be fixed and in accordance with \cite{lee_bayesian_2011} and \cite{zucknick_nonidentical_2015}. 
We set $a_0=2$ and estimate the hyperparameters $\eta_s$ and $\kappa_s$ from the (training) data by fitting a parametric Weibull
model without covariates to the survival data of subgroup $s$.

We perform variable selection using the SSVS approach by \cite{george_variable_1993} in section \ref{sec:ssvs}.
The prior of the regression coefficients $\beta_{s,i}$ in subgroup $s$ conditional on the latent indicator $\gamma_{s,i}$ is defined as a mixture of two normal distributions with small ($\tau^2$) and large ($c^2 \tau^2$) variance 
\[
\beta_{s,i}|\gamma_{s,i} \sim (1-\gamma_{s,i}) \cdot \mathcal{N}(0, \tau^2) + \gamma_{s,i} \cdot \mathcal{N}(0, c^2 \tau^2) \, , \quad i=1,...,p .
\]
The latent indicator variable $\gamma_{s,i}$ indicates the inclusion ($\gamma_{s,i}= 1$) or exclusion (${\gamma_{s,i}= 0}$) of variable $i$ in the model for subgroup $s$.
We assume equal variances for all regression coefficients. 
We set the hyperparameters to the fixed values $\tau=0.0375$ and $c=20$ following \cite{treppmann_integration_2017}.
This choice corresponds to a standard deviation of $c \cdot \tau = 0.75$ and a 95\% probability interval of $[-1.47,1.47]$ for $p(\beta_{s,i}|\gamma_{s,i}= 1)$.

\subsubsection*{Prior on $\boldsymbol{\gamma}$ linking variable and graph selection}

The standard prior for the binary variable selection indicators $\gamma_{s,i}$ is a product of independent Bernoulli distributions as utilized by \cite{treppmann_integration_2017}.
However, this does not consider information from other subgroups and relationships between covariates. 
For this situation, we propose a Markov random field (MRF) prior for the latent variable selection indicators that incorporates information on the relationships among the covariates as described by an undirected graph. 
This prior assumes that neighboring covariates in the graph are more likely to have a common effect and encourages their joint inclusion.
The MRF prior for $\boldsymbol{\gamma}$ given $\boldsymbol{G}$ is defined as
\begin{align}
p(\boldsymbol{\gamma}|\boldsymbol{G}) &= \frac{ \exp( a \boldsymbol{1}_{pS}' \boldsymbol{\gamma}  + b \boldsymbol{\gamma}'\boldsymbol{G} \boldsymbol{\gamma})}{\sum_{\boldsymbol{\gamma} \in \{0,1\}^{pS}} \exp( a \boldsymbol{1}_{pS}' \boldsymbol{\gamma}  + b \boldsymbol{\gamma}'\boldsymbol{G} \boldsymbol{\gamma})} 
\propto \exp( a \boldsymbol{1}_{pS}' \boldsymbol{\gamma}  + b \boldsymbol{\gamma}'\boldsymbol{G} \boldsymbol{\gamma}) , \notag
\end{align}
where $\boldsymbol{\gamma}=(\gamma_{1,1},...,\gamma_{1,p},...,\gamma_{S,1},...,\gamma_{S,p})'$ is a $p S$-dimensional vector of variable inclusion indicators, 
$\boldsymbol{G}$ is a symmetric $(pS \times pS)$ adjacency matrix representation of the graph,
and $a$, $b$ are scalar hyperparameters. 

The hyperparameter $a$ influences the overall variable inclusion probability and controls the sparsity of the model, with smaller values resulting in sparser models. Without loss of generality $a<0$. 
The hyperparameter $b>0$ determines the prior belief in the strength of relatedness between pairs of neighboring variables in the graph and controls the probability of their joint inclusion. 
Higher values of $b$ encourage the selection of variables with neighbors already selected into the model.
The idea becomes more evident by looking at the conditional probability
\[
p(\gamma_{s,i}|\boldsymbol{\gamma}_{-(s,i)},\boldsymbol{G}) = \frac{\exp \left( a \gamma_{s,i} + 2b \gamma_{s,i} \cdot(\sum_{j\neq i} \gamma_{s,j} g_{ss,ij} + \sum_{r\neq s} \gamma_{r,i} g_{rs,ii}) \right)}{1+\exp \left( a +2b \cdot(\sum_{j\neq i} \gamma_{s,j} g_{ss,ij} + \sum_{r\neq s} \gamma_{r,i} g_{rs,ii}) \right)} .
\]
An MRF prior for variable selection has also been used by other authors \cite{li_bayesian_2010, stingo_variable_2011, stingo_incorporating_2011, peterson_joint_2016}. 
However, unlike us, they do not address the problem of borrowing information across subgroups by linking covariates in a graph.

We propose a joint graph with possible edges between all pairs of covariates within each subgroup and edges between the same covariates in different subgroups. 
The elements $g_{rs,ij}$ in the adjacency matrix of the graph $\boldsymbol{G}$ represent the presence ($g_{rs,ij}=1$) or absence ($g_{rs,ij}=0$) of an edge between nodes (genes) $i$ and $j$ in subgroups $r$ and $s$. They can be viewed as latent binary indicator variables for edge inclusion.
The adjacency matrix in the present model is defined as
\[
\boldsymbol{G}= 
\left( \begin{array}{@{}*{4}{c}@{}}
\boldsymbol{G}_{11} & \boldsymbol{G}_{12} & \ldots & \boldsymbol{G}_{1S} \\
\boldsymbol{G}_{12} & \boldsymbol{G}_{22} & \ldots & \boldsymbol{G}_{2S} \\
\vdots & \vdots & \ddots & \vdots \\
\boldsymbol{G}_{1S} & \boldsymbol{G}_{2S} & \ldots & \boldsymbol{G}_{SS} \\
\end{array} \right) .
\]
$\boldsymbol{G}_{ss}=(g_{ss,ij})_{i<j}$ is the matrix of latent edge inclusion indicators within subgroup $s$ 
\[
\boldsymbol{G}_{ss} = 
\begin{pmatrix}
\, 0 \quad & g_{ss,12} & \ldots & g_{ss,1(p-1)} & g_{ss,1p} \\
g_{ss,12} & \, 0 \quad  &  \ddots &  & g_{ss,2p} \\
\vdots & \ddots & \ddots & \ddots & \vdots  \\
g_{ss,1(p-1)} &  & \ddots & \, 0 \quad  & g_{ss,(p-1)p}  \\
g_{ss,1p} &  g_{ss,2p} & \ldots & g_{ss,(p-1)p} &  \, 0 \quad  \\
\end{pmatrix} ,
\]
and $\boldsymbol{G}_{rs}=(g_{rs,ii})_{r<s}$ is the matrix of latent edge inclusion indicators between subgroups $r$ and $s$
\[
\boldsymbol{G}_{rs} = \text{diag}(g_{rs,11},...,g_{rs,pp})  ,
\]
with $r,s=1,...,S$, $r<s$, $i,j=1,...,p$, $i<j$.

\begin{figure}[!htb]
	\centering
  \includegraphics[width=.5\textwidth, trim={0 1.7cm 0 .8cm},clip]{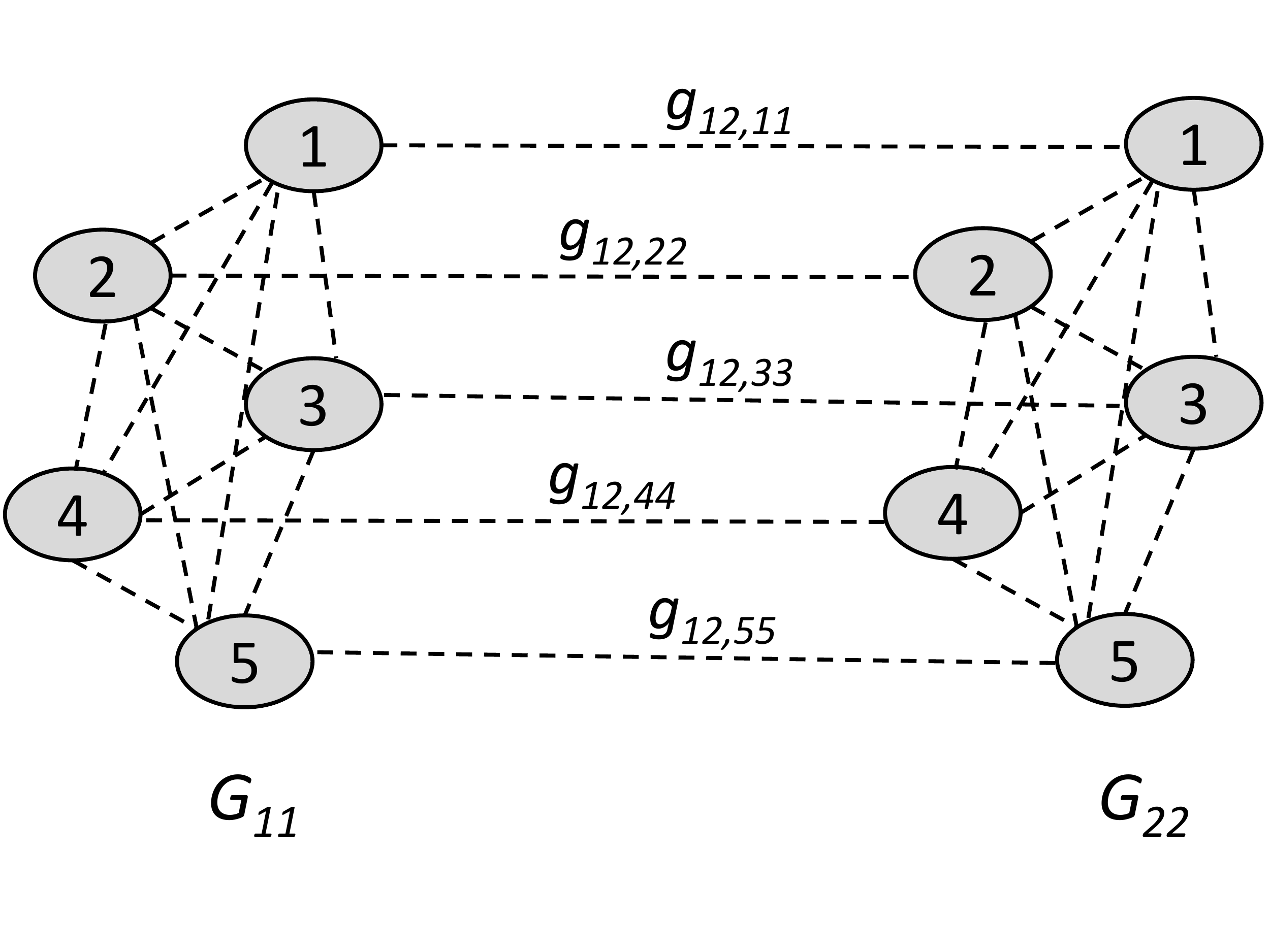}
	\caption{Illustration of the proposed graph for $S=2$ subgroups, each with $p=5$ genomic predictors (nodes). Possible edges between two nodes are marked by dashed lines.}
	\label{fig_bsp_graph}
\end{figure}

Thus, within each subgroup $s$ we assume a standard undirected graph with possible edges between all pairs of genes representing conditional dependencies as in a functional or regulatory pathway. 
Between different subgroups we only allow for relations between the same gene in different subgroups (different genes in different subgroups are assumed to be unconnected). 
This allows sharing information between subgroups and prognostic genes shared by different subgroups have a higher inclusion probability.
To visualize this idea, Figure \ref{fig_bsp_graph} shows an example network consisting of two subgroups, each with five predictors.

\subsubsection*{Graph selection prior on $\boldsymbol{\Omega}$ and $\boldsymbol{G}$}

We infer the unknown graph and precision matrix using the structure learning approach for Gaussian graphical models by \cite{wang_scaling_2015} (section~\ref{sec:bayesnetzw}).
The precision matrix of subgroup $s$ corresponding to subgraph $\boldsymbol{G}_{ss}$ is denoted by ${\boldsymbol{\Omega}_{ss} = (\omega_{ss,ij})_{i<j}}$.
The corresponding prior is defined by 
\[
p(\boldsymbol{\Omega}_{ss}|\boldsymbol{G}_{ss},\nu_0,\nu_1,\lambda) \propto  \prod_{i<j} \mathcal{N}(\omega_{ss,ij}|0,\nu^2_{g_{ss,ij}}) \prod_{i} \text{Exp}(\omega_{ss,ii}|\frac{\lambda}{2}) \mathds{1}_{\{\Omega_s \in \mathcal{M}^{+} \}}, 
\]
with fixed hyperparameters $\nu_0>0$ small, $\nu_1>0$ large and $\lambda>0$. 

We assume the binary edge inclusion indicators within subgroup $s$ ($g_{ss,ij}$) as well as between subgroups $r$ and $s$ ($g_{rs,ii}$) to be independent Bernoulli a priori 
\[
p(\boldsymbol{G}|\pi) \propto \prod_{s} \prod_{i<j} \big[ \pi^{g_{ss,ij}}(1-\pi)^{1-g_{ss,ij}} \big] \cdot \prod_{r<s} \prod_{i} \big[ \pi^{g_{rs,ii}}(1-\pi)^{1-g_{rs,ii}} \big]  ,
\]
with fixed prior probability of edge inclusion $\pi \in (0,1)$.



\subsection{Posterior inference} 

The joint posterior distribution for the set of all parameters $\theta = \{ \boldsymbol{h}, \boldsymbol{\beta}, \boldsymbol{\gamma}, \boldsymbol{G}, \boldsymbol{\Omega} \}$ is proportional to the product of the joint likelihood and the prior distributions of the parameters in all subgroups
\begin{eqnarray*}
\lefteqn{p(\boldsymbol{h}, \boldsymbol{\beta}, \boldsymbol{\gamma}, \boldsymbol{G}, \boldsymbol{\Omega}|\mathfrak{D},\boldsymbol{X})} \\
& \propto &
\prod_{s=1}^S \Big[ L(\mathfrak{D}_s|\boldsymbol{\beta}_s,\boldsymbol{h}_s) \cdot p(\boldsymbol{X}_s|\boldsymbol{\Omega}_{ss}) \Big] 
\cdot 
\prod_{s=1}^S \Big[ p(\boldsymbol{\Omega}_{ss}|\boldsymbol{G}_{ss}) \cdot p(\boldsymbol{G}) \cdot p(\boldsymbol{\gamma}|\boldsymbol{G}) \cdot 
\prod_{i=1}^{p} p(\beta_{s,i}|\gamma_{s,i}) \cdot 
\prod_{g=1}^{J_s} p(h_{s,g}|\boldsymbol{\beta}_s) \Big] . 
\end{eqnarray*}

\subsubsection{Markov chain Monte Carlo sampling} \label{sec:MethMCMC}

Markov Chain Monte Carlo (MCMC) simulations are required to obtain a posterior sample of the parameters. The different parameters are updated iteratively according to their conditional posterior distributions using a Gibbs sampler.
A brief outline of the MCMC sampling scheme is given in the following. More details are provided in Supplementary Materials.

\begin{enumerate}
\item 
For subgroup $s=1,...,S$ update $\boldsymbol{\Omega}_{ss}$ with the block Gibbs sampler proposed by \cite{wang_scaling_2015}.
\item 
Update all elements in $\boldsymbol{G}$ iteratively with Gibbs sampler from the conditional distributions ${p(g_{ss,ij} =1| \boldsymbol{G}_{-ss,ij}, \omega_{ss,ij}, \boldsymbol{\gamma})}$ as well as ${p(g_{rs,ii}=1 | \boldsymbol{G}_{-rs,ii}, \boldsymbol{\gamma})}$, where $\boldsymbol{G}_{-rs,ii}$ ($\boldsymbol{G}_{-ss,ij}$) denotes all elements in $\boldsymbol{G}$ except for $g_{rs,ii}$ ($g_{ss,ij}$).
\item
Update all elements in $\boldsymbol{\gamma}$ iteratively with Gibbs sampler from the conditional distributions ${p(\gamma_{s,i} =1 | \boldsymbol{\gamma}_{-s,i}, \boldsymbol{G}, \beta_{s,i})}$, where $\boldsymbol{\gamma}_{-s,i}$ denotes all elements in $\boldsymbol{\gamma}$ except for $\gamma_{s,i}$.
\item
Update $\beta_{s,i}$ from the 
conditional distribution $p(\beta_{s,i}|\boldsymbol{\beta}_{s,-i}, \boldsymbol{\gamma}_s, \boldsymbol{h}_s, \mathfrak{D}_s)$, ${s=1,...,S}$, ${i=1,...,p}$, using a random walk Metropolis-Hastings algorithm with adaptive jumping rule as proposed by \cite{lee_bayesian_2011}. $\boldsymbol{\beta}_{s,-i}$ includes all elements in $\boldsymbol{\beta}_{s}$ except for $\beta_{s,i}$.
\item
The 
conditional distribution $p(h_{s,g}|\boldsymbol{h}_{s,-g}, \boldsymbol{\beta}_s, \boldsymbol{\gamma}_s, \mathfrak{D}_s)$ for the update of $h_{s,g}$ can be well approximated by the gamma distribution 
\[
h_{s,g}|\boldsymbol{h}_{s,-g}, \boldsymbol{\beta}_s, \boldsymbol{\gamma}_s, \mathfrak{D}_s 
 \overset{\text{approx.}}{\sim}  
\mathcal{G} \Big(a_0 (H^\ast(c_{s,g})-H^\ast(c_{s,g-1})) + d_{s,g}, a_0 + \mspace{-20mu} \sum_{k \in \mathcal{R}_{s,g}-\mathcal{D}_{s,g}} \mspace{-20mu} \exp(\boldsymbol{\beta}_s' \boldsymbol{x}_{s,k}) \Big),
\]
where $d_{s,g}$ is the number of events in interval $g$ for subgroup $s$ and $\boldsymbol{h}_{s,-g}$ denotes the vector $\boldsymbol{h}_{s}$ without the $g$-th element, $g=1,...,J_s$, $s=1,...,S$ \cite[chapter 3.2.2]{ibrahim_bayesian_2005}.
\end{enumerate}

Starting with an arbitrary set of initial values for the parameters, the MCMC algorithm runs with a reasonably large number of iterations to obtain a representative sample from the posterior distribution. 
All subsequent results are based on single MCMC chains, each with 20 000 iterations in total and a burn-in period of 10 000 iterations.
As starting values we choose an empty model with:
\begin{quote}
$\boldsymbol{G}^{(0)}=\boldsymbol{0}_{pS\times pS}$ \\ 	\vspace{0.02cm} \\
	$\boldsymbol{\Sigma}_s^{(0)}=\boldsymbol{I}_{p\times p}$ and $\boldsymbol{\Omega}_{ss}^{(0)}=(\boldsymbol{\Sigma}_s^{(0)})^{-1}$ for $s=1,...,S$ \\
	\vspace{0.02cm} \\
	$\boldsymbol{\gamma}_s^{(0)}=(0,...,0)'$ for $s=1,...,S$ \\ 
	\vspace{0.02cm} \\
	$\beta_{s,i}^{(0)} \sim \mathcal{U}[-0.02, 0.02]$ for $i=1,...,p$, $s=1,...,S$ \\ 
	\vspace{0.02cm} \\
	$h_{s,g}^{(0)} \sim \mathcal{G}(1,1)$ for $s=1,...,S$, $g=1,...,J_s$. \\
\end{quote}

We assessed convergence of each MCMC chain by looking at autocorrelations, trace plots and running mean plots of the regression coefficients. 
In addition, we ran several independent MCMC chains with different starting values to ensure that the chains and burn-in period were long enough to reach (approximate) convergence.

\subsubsection{Posterior estimation and variable selection}

We report the results of the Cox models in terms of marginal and conditional posterior means and standard deviations of the estimated regression coefficients, as well as posterior selection probabilities. After removal of the burn-in samples, the remaining MCMC samples serve as draws from the posterior distribution to calculate the empirical estimates. 
These estimates are then averaged across all training sets for each variable separately.

The strategy for variable selection follows \cite{treppmann_integration_2017}. First, the mean model size $m^\ast$ is computed as the average number of included variables across all MCMC iterations after the burn-in. Then the $m^\ast$ variables with the highest posterior selection probability 
are considered as the most important variables and selected in the final model.

\subsubsection{Prediction}

We use training data for model fitting and posterior estimation and test data to assess model performance.
We evaluate the prediction performance of the Cox models by 
the integrated Brier score.

The expected Brier score can be interpreted as a mean square error of prediction.
It measures the inaccuracy by comparing the estimated survival probability $\hat{S}(t|\boldsymbol{x}_m)$ of a patient $m$, $m=1,..,n$, with the observed survival status $\mathds{1}(\tilde{t}_m > t)$ 
\[
\widehat{\textsl{BS}}(t)= \frac{1}{n} \sum_{m=1}^n \hat{w}_m(t) \cdot \left( \mathds{1}(\tilde{t}_m > t) -\hat{S}(t|\boldsymbol{x}_m) \right)^2 
\]
and the squared residuals are weighted using inverse probability of censoring weights
\[
\hat{w}_m(t) = \frac{ \mathds{1}(\tilde{t}_m \leq t) \delta_m}{\hat{C}(\tilde{t}_m)} + \frac{\mathds{1}(\tilde{t}_m > t) }{\hat{C}(t)} 
\]
to adjust for the bias caused by the presence of censoring in the data. 
$\hat{C}(t)$ is the Kaplan-Meier estimator of the censoring times \cite{schumacher_assessment_2007, binder_overview_2011}.


The predictive performance of competing survival models can be compared by plotting the Brier score over time (prediction error curves).
Alternatively, prediction error curves can be summarized in one value with the integrated Brier score as a measure of inaccuracy over a time interval rather than at single time points \cite{graf_assessment_1999}
\[
\textsl{IBS}(t^\ast)= \frac{1}{t^\ast} \int_0^{t^\ast} \textsl{BS}(t) \text{d}t , \quad t^\ast > 0 .
\]

\subsubsection{Median Probability Model and Bayesian Model Averaging}

For the calculation of the prediction error, we account for the uncertainty in model selection by two different approaches: the Median Probability Model (MPM) \cite{barbieri2004} and an approximation to Bayesian Model Averaging (BMA) \cite{Hoeting_Bayesian_1999}.
After removal of the burn-in samples, we compute the Brier score over the ``best'' selected models.
According to the BMA approach we choose the top 100 models with the largest 
log-likelihood values to obtain the marginal posterior means of the regression coefficients, which in turn are required for the risk score.
For the MPM approach we select all covariates with a mean posterior selection probability larger than 0.5. For these variables we calculate the marginal posterior means of the regression coefficients and the corresponding risk score.



\section{Simulation study}
\label{sec:simstudy}

In section~\ref{sec:simres1} we compare the performance of our proposed model, referred to as \textsl{CoxBVS-SL} (for Cox model with Bayesian Variable Selection and Structure Learning, as an extension of the model by \cite{treppmann_integration_2017}), to a standard subgroup model and a combined model.
The combined model pools data from all subgroups and treats them as one homogeneous cohort, whereas the subgroup model only uses information in the subgroup of interest and ignores the other subgroups. 
Both standard approaches follow the Bayesian Cox model proposed by \cite{treppmann_integration_2017} with stochastic search variable selection and independent Bernoulli priors for the variable inclusion indicators $\boldsymbol{\gamma}$. 

The priors for variable selection and structure learning are specified as follows.
We set the hyperparameter of the Bernoulli distribution to ${\pi=0.02}$, 
matching the prior probability of variable inclusion in the MRF prior of the CoxBVS-SL model. 
Based on a sensitivity analysis, we choose the hyperparameters of the MRF prior as ${a=-4}$ and ${b=1}$.
When the graph $\boldsymbol{G}$ contains no edges or ${b=0}$ then the prior variable inclusion probability is $\frac{exp(a)}{(1+exp(a))}\approx 0.018$.
This probability increases when ${b>0}$ is combined with a nonempty graph.
The remaining hyperparameters for $\boldsymbol{G}$ and $\boldsymbol{\Omega}_{ss}$ are chosen as $\nu_0 = 0.1, \nu_1 = 10, \lambda=1$ 
and $\pi=2/(p-1)$, 
following the recommendations in \cite{wang_scaling_2015} and \cite{peterson_joint_2016}.

We examine varying numbers of genomic covariates $p$ and sample sizes $n$, with a focus on small sample sizes relative to the number of variables which is characteristic for gene expression data.
We standardize the genomic covariates before model fitting and evaluation to have zero mean and unit variance.
Parameters of the training data (mean and standard deviation of each variable) are used to scale the training and test data.
For the standard subgroup model and the proposed model we standardize each subgroup separately, whereas for the combined model we pool training data of all subgroups. 

For Bayesian inference, typically one training data set is used for posterior estimation and an independent test data set for model evaluation. 
However, results have shown some variation due to the data draw. Therefore, in the following, simulation of training and test data is repeated ten times for each simulation scenario.

In section~\ref{sec:simres2} we use two different hyperparameters $b$ for the subgraphs $\boldsymbol{G}_{ss}$, $s=1,2$ and $\boldsymbol{G}_{12}$ in the MRF prior of the CoxBVS-SL model and compare the prediction performance with the \textit{Sub-struct} model. In the latter $\boldsymbol{G}_{12}$ is an empty graph and only information of $\boldsymbol{G}_{ss}$ is included in the MRF prior. 
We use the same training and test data as in section~\ref{sec:simres1} but only consider simulation scenarios with $p=100$.

\subsection{Data simulation}

Training and test data each consisting of $n$ samples and $p$ genomic covariates are simulated from the same distribution 
as described in the following.
We consider two subgroups that differ only in their relationship between genomic covariates and survival endpoint ($\boldsymbol{\beta}_s$, $s=1,2$), and in the parameters for the simulation of survival data. 
We generate gene expression data from the same multivariate normal distribution with mean vector $\boldsymbol{0}$ and covariance matrix $\boldsymbol{\Sigma}$. 
The corresponding precision matrix $\boldsymbol{\Omega} = \boldsymbol{\Sigma}^{-1}$ is defined such that the variance of each gene is 1 and partial correlations exist only between the first nine prognostic genes. 
Within the three blocks of prognostic genes determined by the same effect (gene 1 to 3, gene 4 to 6, and gene 7 to 9) we assume pairwise partial correlations of 0.5.
All remaining genes are assumed to be uncorrelated.

We simulate survival data from a Weibull distribution according to \cite{bender_generating_2005}, with scale $\eta_s$ and shape $\kappa_s$ parameters estimated from two real gene expression cancer cohorts. 
Therefore, we compute survival probabilities at 3 and 5 years using the Kaplan-Meier estimator for both cohorts separately. 
The corresponding probabilities are 57\% and 75\% for 3-years survival, and 42\% and 62\% for 5-years survival, respectively. 
Individual event times for subgroup $s$ are simulated as
\[
	T_{s} \sim \left(- \frac{\log(U)}{\eta_{s} \exp(\boldsymbol{x}_{s} \boldsymbol{\beta}_{s})}\right)^{1/\kappa_{s}} , \quad U \sim \mathcal{U}[0,1],
\]
with true effects $\boldsymbol{\beta}_{s} \in \mathds{R}^{p}$, $s=1,2$. 
We randomly draw noninformative censoring times $C_s$ from a Weibull distribution with the same parameters as for the event times, resulting in approximately 50\% censoring rates in both subgroups.
The individual observed event indicators and times until an event or censoring are defined as $\delta_s = \mathds{1}(T_s \leq C_s)$ and  $\widetilde{T_s}=\min(T_s,C_s)$, $s=1,2$.

We choose the true effects of the genomic covariates on survival outcome as stated in Table~\ref{tab:simeffects}.
Genes 1, 2, 3 and 7, 8, 9 are subgroup-specific, while genes 4, 5 and 6 have the same effect in both subgroups. All remaining genes represent noise and have no effect in both subgroups. 
\begin{table}[!htb]
	\small
	\centering
	\caption{True effects in both subgroups for the simulation of survival outcome.}
\begin{tabular}{c|cccccccccccc}
  & \multicolumn{12}{c}{Gene} \\
	& 1 & 2 & 3 & 4 & 5 & 6 & 7 & 8 & 9 & 10 & $\ldots$ & $p$ \\
\hline
$\boldsymbol{\beta}_1$ & 1 & 1 & 1 & -1 &-1 & -1 & 0 & 0 & 0 & 0 & $\ldots$ & 0 \\
$\boldsymbol{\beta}_2$ & 0 & 0 & 0 & -1 &-1 & -1 & 1 & 1 & 1 & 0 & $\ldots$ & 0 \\
\end{tabular}
\label{tab:simeffects}
\end{table}

\subsection{Simulation results I}
\label{sec:simres1}

We consider three low-dimensional settings with $p=20$ genes and $n=50,75,100$ samples in each subgroup, as well as five high-dimensional settings with $p=100$ and sample sizes $n=50, 75, 100,$ $150$.
We also tested $p=100$ and $n=125$, but as expected, the results always lay between the results for $n=100$ and $n=150$. For this reason, they are not shown here.
We compare our proposed model (\textsl{CoxBVS-SL}) to the standard subgroup model (\textsl{Subgroup}) and the standard combined or pooled model (\textsl{Pooled}) regarding variable selection accuracy and prediction performance. 

Posterior selection probabilities for each gene are computed based on all iterations after the burn-in and averaged across all training data sets.
The resulting mean posterior selection probabilities of the first nine genes in subgroup 1 are depicted in Figure~\ref{fig:BayesSim1PPI1} (and in Supplementary Figure~\ref{fig:BayesSim1PPI2} for subgroup 2).
Across all simulation scenarios, the CoxBVS-SL model has more power for the selection of prognostic genes compared to the two standard approaches, and at the same time, does not erroneously select noise genes (false positives) as the Pooled model.
As expected, with larger $n$, power and accuracy in variable selection increase for both, the CoxBVS-SL and the Subgroup model.
The Pooled model only correctly identifies the joint effects of genes 4, 5 and 6 
but fails to detect subgroup-specific effects.

\begin{figure}[!htb] 
\begin{knitrout}
\definecolor{shadecolor}{rgb}{0.969, 0.969, 0.969}\color{fgcolor}

{\centering \includegraphics[width=\linewidth]{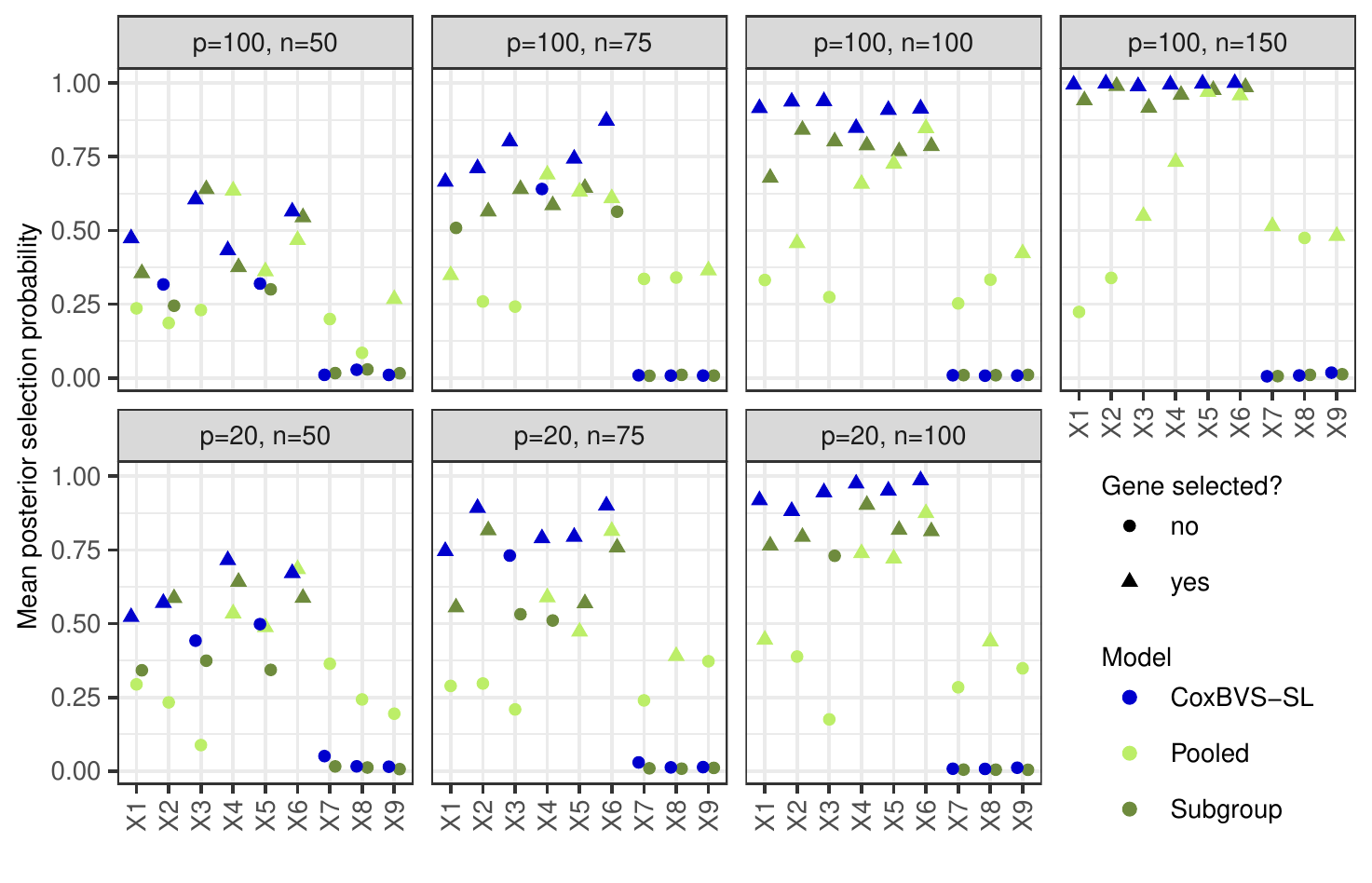} 

}

\end{knitrout}
\caption{Mean posterior selection probabilities of the first nine genes in subgroup~1 (averaged across the ten training sets). The colors represent the different models and the plot symbol indicates whether a gene is selected on average or not.} 
\label{fig:BayesSim1PPI1}
\end{figure}

Posterior estimates of the regression coefficients $\hat{\beta}_j$ of the first nine genes in subgroup 1 are shown in Figure~\ref{fig:BayesSim1Betas2} for conditional posterior means (conditional on $\gamma=1$) and in Supplementary Figure~\ref{fig:BayesSim1Betas1} for marginal posterior means (independent of $\gamma$), both along with standard deviations.
The corresponding results for subgroup 2 are depicted in Supplementary Figures~\ref{fig:BayesSim1Betas2s2} and~\ref{fig:BayesSim1Betas1s2}.
For $n<100$ the conditional posterior means of the prognostic genes are less shrunk than the marginal posterior means.
Results of the CoxBVS-SL model and the Subgroup model are very similar, whereas the Pooled model averages effects across subgroups leading to biased subgroup-specific effects and more false positives.
Surprisingly, the joint effects of genes 4, 5 and 6 
are also more precisely estimated (less shrunk) by CoxBVS-SL and Subgroup compared to Pooled.

\begin{figure}[!hbt] 
\begin{knitrout}
\definecolor{shadecolor}{rgb}{0.969, 0.969, 0.969}\color{fgcolor}

{\centering \includegraphics[width=\linewidth]{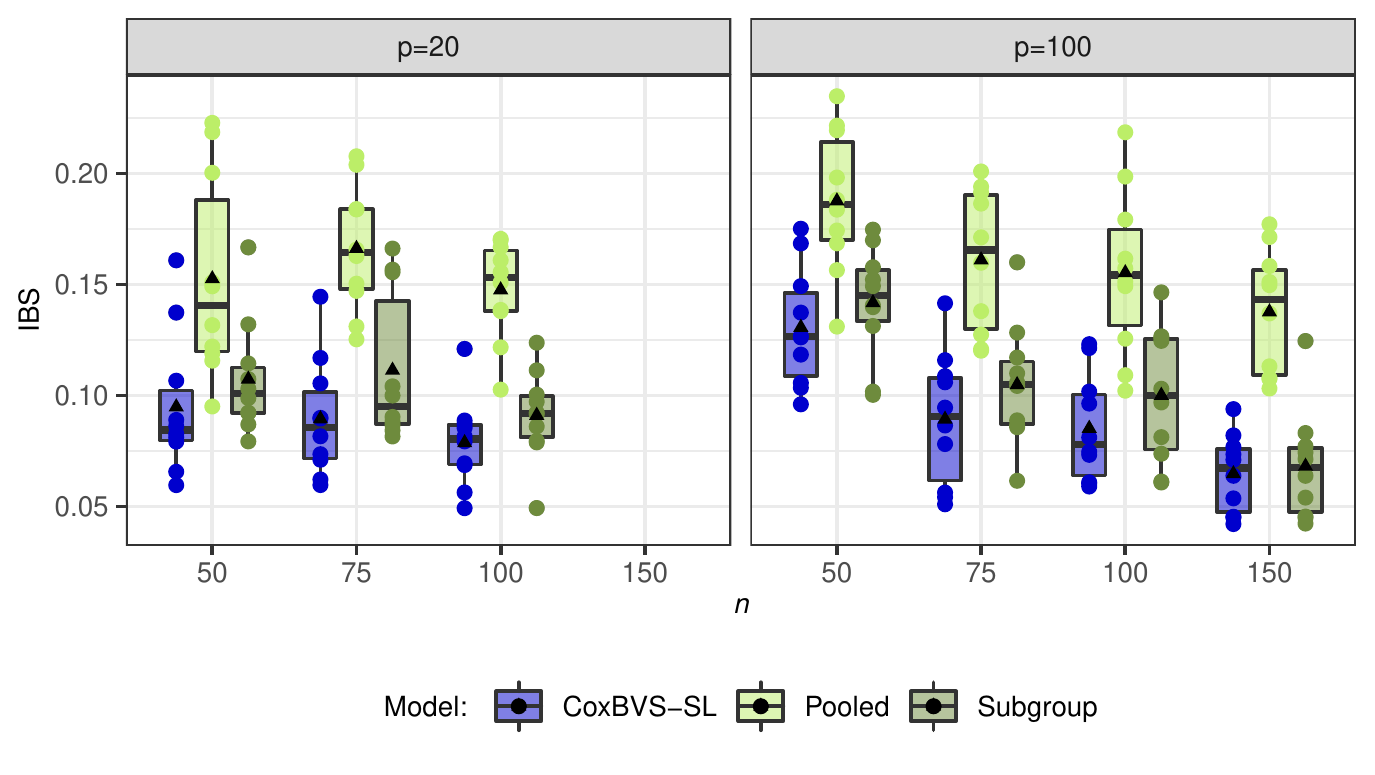} 

}

\end{knitrout}
\caption{Integrated Brier Scores (IBS) across all ten test sets for subroup~1 (IBS based on the Median Probability Model). The black triangle within each boxplot represents the mean value.} 
\label{fig:BayesSim1IBSMPM}
\end{figure}

We assess prediction performance by the integrated Brier Score (IBS), computed based on the Median Probability Model (MPM, Figure~\ref{fig:BayesSim1IBSMPM} for subroup 1 and Supplementary Figure~\ref{fig:BayesSim1IBSMPMs2} for subgroup 2) and the Bayesian Model Averaging (BMA, Supplementary Figure~\ref{fig:BayesSim1IBSBMA} for subroup 1 and Supplementary Figure~\ref{fig:BayesSim1IBSBMAs2} for subgroup 2).
The Pooled model has the worst prediction accuracy. 
In the case of MPM, CoxBVS-SL performs clearly better than Subgroup, for BMA both models are competitive.

\begin{figure}[!htb] 
\begin{knitrout}
\definecolor{shadecolor}{rgb}{0.969, 0.969, 0.969}\color{fgcolor}

{\centering \includegraphics[width=\linewidth]{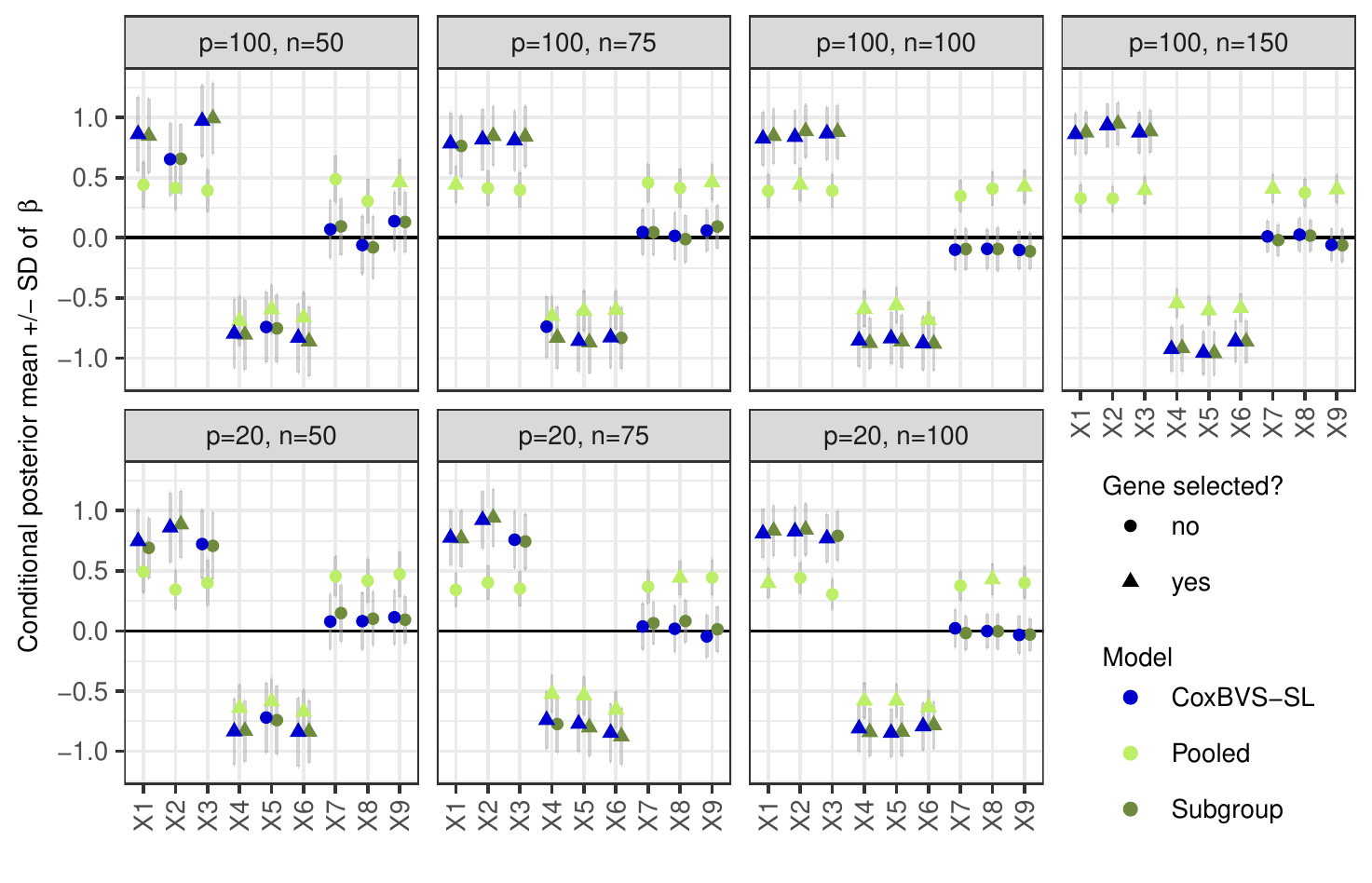} 

}

\end{knitrout}
\caption{Conditional posterior means (conditional on $\gamma=1$) and standard deviations (SD) of the regression coefficients of the first nine genes in subgroup~1 (averaged across the ten training sets). 
} 
\label{fig:BayesSim1Betas2}
\end{figure}

\clearpage

Inference of the graph showed relatively high accuracy for learning the conditional dependence structure among genes within
subgroups and for detecting joint effects across different subgroups.
The block correlation structure between the prognostic genes within each subgroup is correctly estimated by the precision matrix and the subgraph $\boldsymbol{G}_{ss}$, $s=1,2$ in the CoxBVS-SL model (see Supplementary Figure~\ref{fig:BayesSim1Gss}).
Inference of the subgraph $\boldsymbol{G}_{12}$ linking both subgroups improves with increasing sample size. 
The corresponding marginal posterior edge inclusion probabilities of the prognostic genes with joint effects (genes 4, 5 and 6) are larger than for the remaining genes, which becomes more evident for increasing $n$ (see Supplementary Figure~\ref{fig:BayesSim1G12}).
Findings support the assumption that incorporating network information into variable selection may increase power to detect associations with the survival outcome and improve prediction accuracy. 


\subsection{Simulation results II}
\label{sec:simres2}

Next, we study the effect of two different hyperparameters $b$ in the MRF prior of the CoxBVS-SL model with respect to variable selection and prediction performance.
The new hyperparameter $b_1=1$ corresponds to the subgraphs $\boldsymbol{G}_{ss}$, $s=1,2$ within each subgroup and $b_2 =1,1.5,2,2.5,3$ to the subgraph $\boldsymbol{G}_{12}$ linking both subgroups.
By choosing a larger value for $b_2$, we give $\boldsymbol{G}_{12}$ more weight in the MRF prior and thus, increase the prior variable inclusion probability for genes being simultaneously selected in both subgroups and having a link in $\boldsymbol{G}_{12}$.

We compare the results of CoxBVS-SL with varying $b_2$ to the results of the \textsl{Sub-struct} model where $b_2=0$ 
and only information of $\boldsymbol{G}_{ss}$, $s=1,2$ is included in the MRF prior. 
In this comparison we investigate how much information is added by $\boldsymbol{G}_{12}$ over $\boldsymbol{G}_{ss}$.
For the other hyperparameters we use the same values as in the previous section. 
We apply all models to the same training and test data sets as in section~\ref{sec:simres1} but only consider simulation scenarios with $p=100$ and $n=50,75,100,125,150$.

\begin{figure}[!htb] 
\begin{knitrout}
\definecolor{shadecolor}{rgb}{0.969, 0.969, 0.969}\color{fgcolor}

{\centering \includegraphics[width=\linewidth]{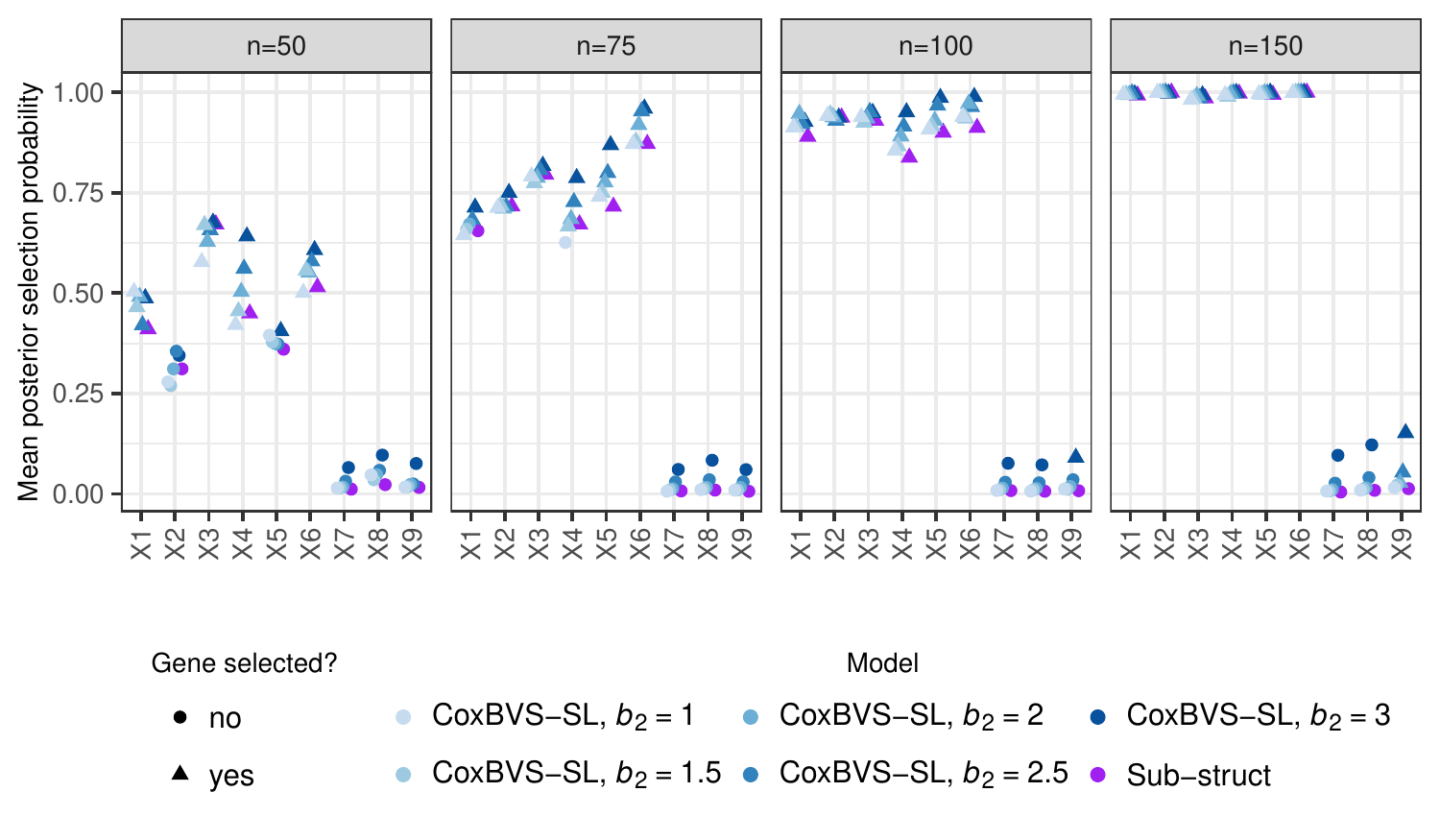} 

}

\end{knitrout}
\caption{Mean posterior selection probabilities (averaged across the ten training sets) of the first nine genes in subgroup 1.} 
\label{fig:BayesSim2PPI}
\end{figure}

Figure~\ref{fig:BayesSim2PPI} shows the mean posterior selection probabilities of the first nine genes in subgroup 1 (subgroup 2 is presented in Supplementary Figure~\ref{fig:BayesSim2PPIs2}).
The results of Sub-struct are similar to CoxBVS-SL with $b_2=1$. 
Increasing values of $b_2$ lead to larger posterior variable inclusion probabilities, however, not only for the prognostic genes (see genes 7, 8 and 9 in subgroup 1). 
This means more power for the correct identification of prognostic genes when $n\leq p$, but on the other hand, a tendency towards more false positives.

\begin{figure}[!hbt] 
\begin{knitrout}
\definecolor{shadecolor}{rgb}{0.969, 0.969, 0.969}\color{fgcolor}

{\centering \includegraphics[width=\linewidth]{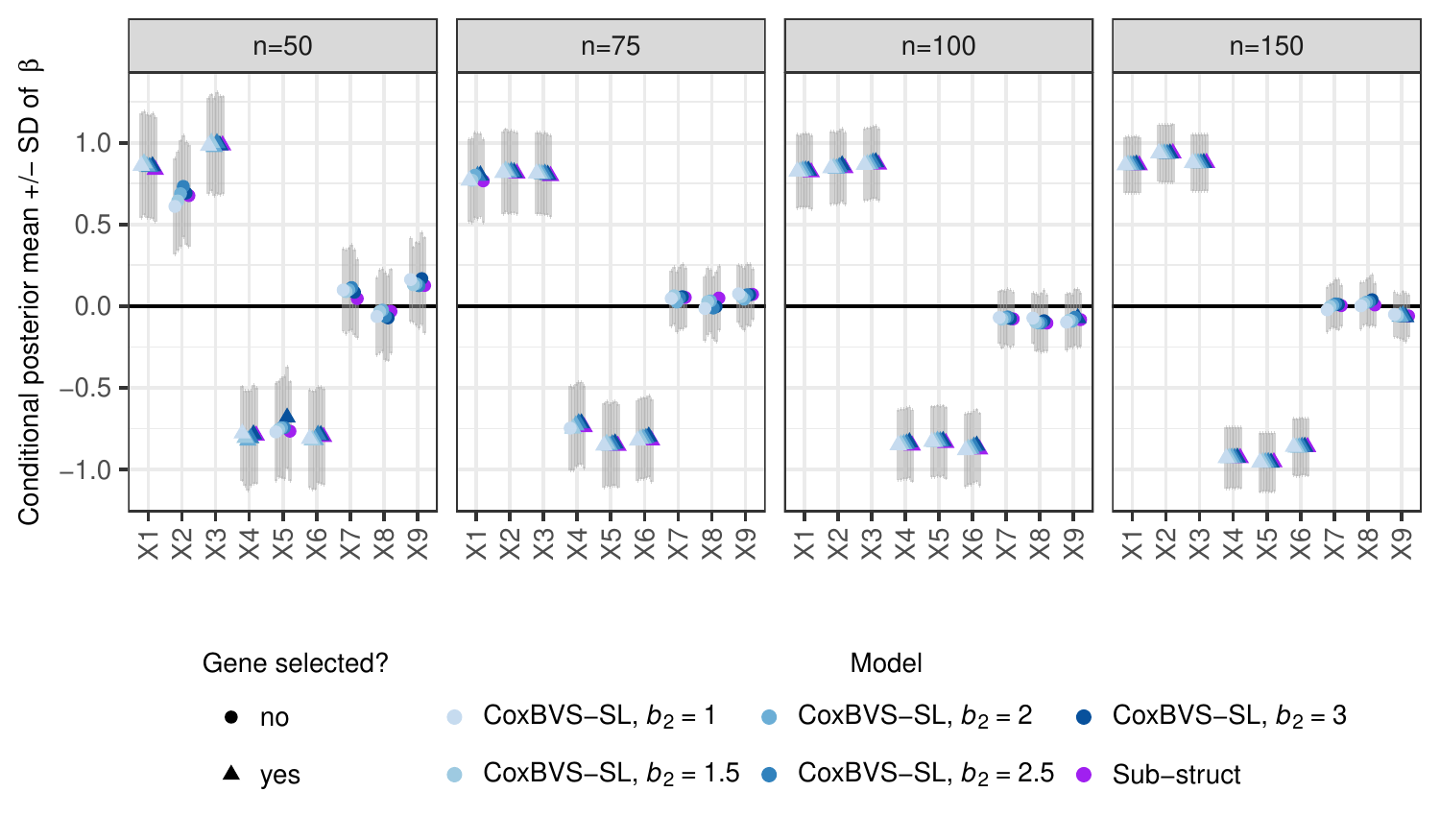} 

}

\end{knitrout}
  \caption{Conditional posterior means (conditional on $\gamma=1$) and standard deviations (SD) of the regression coefficients of the first nine genes in subgroup 1 (averaged across the ten training sets). 
  } 
\label{fig:BayesSim2Betas2}
\end{figure}

Posterior estimates of the regression coefficients $\hat{\beta}_j$ are very similar for all models.
Figure~\ref{fig:BayesSim2Betas2} shows the conditional posterior means (conditional on ${\gamma=1}$) and Supplementary Figure~\ref{fig:BayesSim2Betas1} the marginal posterior means (independent of $\gamma$) along with standard deviations of the first nine genes in subgroup 1.
The corresponding results of subgroup 2 are depicted in Supplementary Figures~\ref{fig:BayesSim2Betas2s2} and~\ref{fig:BayesSim2Betas1s2}.

\begin{figure}[!htb]
\begin{knitrout}
\definecolor{shadecolor}{rgb}{0.969, 0.969, 0.969}\color{fgcolor}

{\centering \includegraphics[width=\linewidth]{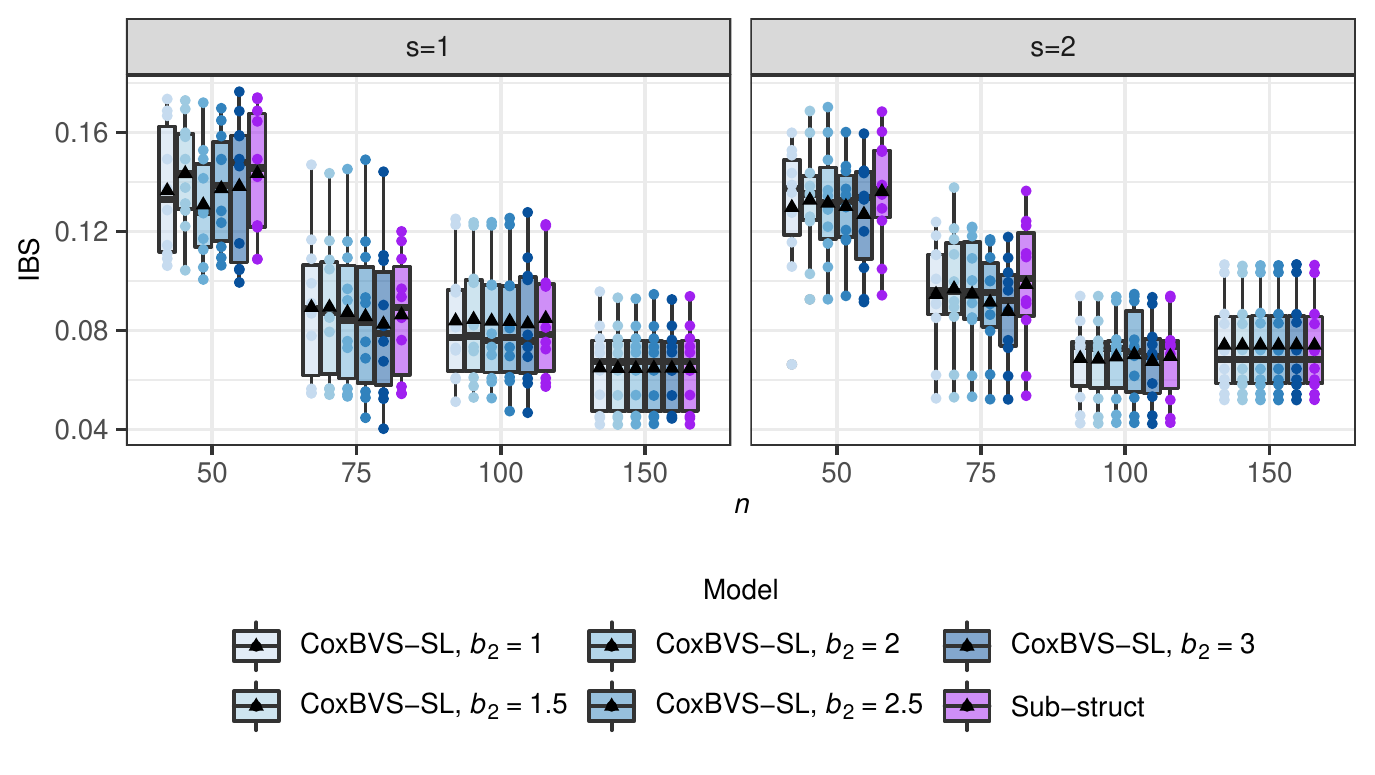} 

}

\end{knitrout}
\caption{Integrated Brier Scores (IBS) across all ten test sets for subroup 1 (left) and 2 (right) (based on the Median Probability Model). The black triangle within each boxplot represents the mean value.} \label{fig:BayesSim2IBSMPM}
\end{figure}

We assess prediction performance in terms of the integrated Brier Score (IBS), computed based on the Median Probability Model (Figure~\ref{fig:BayesSim2IBSMPM}) and the Bayesian Model Averaging (Supplementary Figure~\ref{fig:BayesSim2IBSBMA}).
Larger values of $b_2$ tend to lead to a slightly better prediction performance of CoxBVS-SL compared to Sub-struct when $n<p$.
When the sample size is large, the prediction accuracy of all models is similarly good.

Supplementary Figure~\ref{fig:BayesSim2G12} compares the results of the subgraph $\boldsymbol{G}_{12}$ for varying $b_2$ in CoxBVS-SL. 
For larger values of $b_2$ the marginal posterior edge inclusion probabilities of the prognostic genes with joint effects (genes 4, 5 and 6) increase, as expected, since they are given a higher weight in the prior.
However, when $b_2=3$ we also notice a minor increase of the marginal posterior edge inclusion probabilities of the other six prognostic genes with subgroup-specific effects.


\section{Case study based on Glioblastoma protein expression data}
\label{sec:gbd}

In this section we compare CoxBVS-SL with varying $b_2$ to both standard models, Pooled and Subgroup. 
We use the Glioblastoma protein expression data from \cite{peterson_joint_2016}, comprising 212 samples with survival data (159 events) and $p=187$ proteins.
For reasons of computation time, we use only $p=20$ proteins and standardize the protein expression data as described in section~\ref{sec:simstudy}.
In contrast to the simulated gene expression data in the previous section, we have real correlations between all covariates and the data is not drawn from a multivariate normal distribution.
We still simulate the relationship between proteins and survival outcome by choosing artificial effects and simulating the survival data from a Weibull distribution.
We randomly divide the complete data set into two equally large subsets to obtain two subgroups. 

For the survival endpoint we simulate the event times $T_s$ and censoring times $C_s$, respectively, in subgroup $s$ from a Weibull distribution 
with scale and shape parameters estimated by the Kaplan-Meier estimator of the true event and censoring times, respectively, in the specific subgroup.
The individual observed event indicators and survival times until an event or censoring are defined as $\delta_s = \mathds{1}(T_s \leq C_s)$ and  $t_s=\min(T_s,C_s)$, resulting in approximately 42\% censoring rates in both subgroups.
The effects in subgroup $s=1$ and $s=2$ that we assume for the simulation of survival data are depicted in Table~\ref{tab:GBeffects}.
\begin{table}[htb]
\centering
\footnotesize
\caption{Simulated effects in both subgroups. Groups of proteins with the same effect are defined by different phosphorylation sites (or isoforms) of the same protein.}
	\begin{tabular}{lcc}
\textbf{Protein}	& $\boldsymbol{\beta}_1$ & $\boldsymbol{\beta}_2$ \\
\hline
Akt & 2 & 0\\ 
Akt\_pS473 & 2 & 0\\ 
Akt\_pT308 & 2 & 0\\ \arrayrulecolor{gray}\hline
EGFR & 0 & 2\\ 
EGFR\_pY1068 & 0 & 2\\ 
EGFR\_pY1173 & 0 & 2\\  \arrayrulecolor{gray}\hline
AMPK\_alpha & -1.5 & 1.5\\ 
Annexin.1 & 1.5 & -1.5\\  \arrayrulecolor{gray}\hline
GSK3.alpha.beta & -2 & -2\\ 
GSK3.alpha.beta\_pS21\_S9 & -2 & -2\\ 
GSK3\_pS9 & -2 & -2\\  \arrayrulecolor{gray}\hline
X14.3.3\_beta & 0 & 0\\ 
X14.3.3\_epsilon & 0 & 0\\ 
X14.3.3\_zeta & 0 & 0\\ 
X4E.BP1 & 0 & 0\\ 
X4E.BP1\_pS65 & 0 & 0\\ 
X4E.BP1\_pT37T46 & 0 & 0\\ 
X4E.BP1\_pT70 & 0 & 0\\ 
X53BP1 & 0 & 0\\ 
A.Raf\_pS299 & 0 & 0\\
\end{tabular} 
\normalsize
\label{tab:GBeffects}
\end{table}

We repeatedly randomly split the complete data into training (with proportion 0.8) 
and test sets, stratified by subgroup and event indicator.
In total, we generate ten training data sets for model fitting and ten test data sets for evaluation of the prediction
performance.

We choose the hyperparameters in accordance with the case study in \cite{peterson_joint_2016} as follows.
For the two standard models a prior probability of variable inclusion of 0.2 is assumed. 
In the CoxBVS-SL model we set the hyperparameters 
of the precision matrix and graph to $\nu_0=0.6, \nu_1=360, \lambda=1$ and $\pi=2/(p-1)$.
The hyperparameters of the MRF prior are $a=-1.75, b=0.5$ 
and as in section~\ref{sec:simres2}, we tried out two different values for $b$: 
$b_1=0.5$ and $b_2=1,1.25,1.5,1.75,2,2.25,2.5,2.75,3$, or $b_1=1,1.5,2,2.5,3$ and $b_2=0.5$. 


\subsection{Results of the case study}

\begin{figure}[!htb]
\begin{knitrout}
\definecolor{shadecolor}{rgb}{0.969, 0.969, 0.969}\color{fgcolor}

{\centering \includegraphics[width=\linewidth]{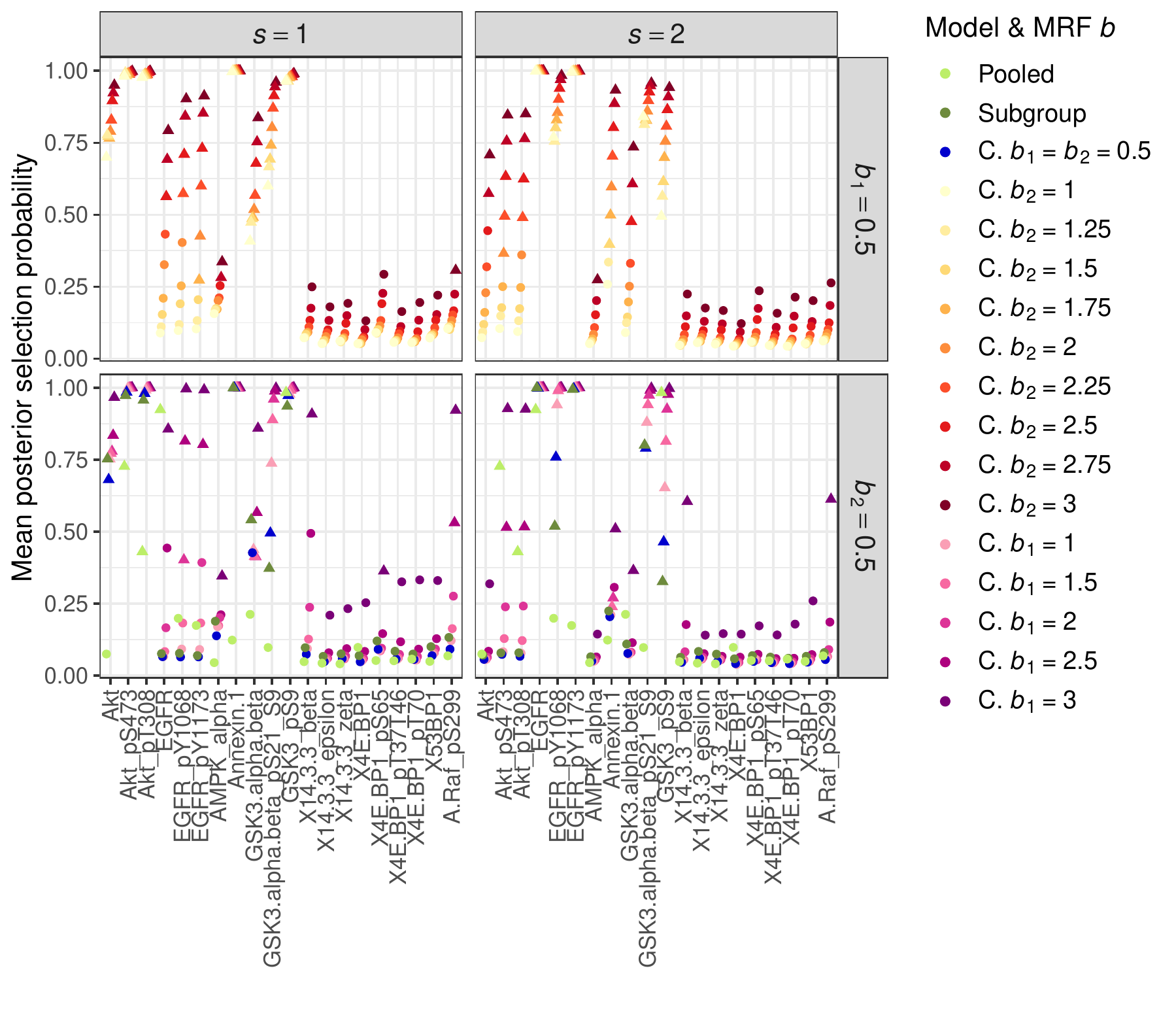} 

}

\end{knitrout}
\caption{Mean posterior selection probabilities of all 20 proteins in both subgroups (averaged across all training sets). The different colors represent the models or parameter values of $b_1$ and $b_2$ in CoxBVS-SL (abbreviated by "C."). The plot symbol indicates whether a protein is selected (triangle) or not (circular point).} 
\label{fig:GBPPI}
\end{figure}

When either $b_1$ or $b_2$ increases the mean posterior selection probabilities of all proteins increase too (Figure~\ref{fig:GBPPI}).
The Subgroup and CoxBVS-SL model with ${b_1 = b_2 = 0.5}$ perform similarly.
They correctly identify the subgroup-specific effects of the first six proteins and do not falsely select any noise proteins. 
Interestingly, the effects of proteins AMPK and Annexin (ID 7 and 8), going in opposite directions for both subgroups, as well as the joint effects of proteins GSK3 are not all identified. 
There are a few false negatives.
The Pooled model, in contrast, shows a clear bias for the subgroup-specific and opposite effects. 
The effects are averaged across both subgroups, which also becomes evident when looking at the posterior estimates of the coefficients, for the conditional posterior means in Figure~\ref{fig:GBcondBetas} and for the marginal posterior means in Supplementary Figure~\ref{fig:GBmargBetas}.
The results of the Subgroup and CoxBVS-SL model are similar. 
In particular, the posterior means of the noise proteins 
are close to 0, also for large values of $b_1$ or $b_2$.

\begin{figure}[!htb]
\begin{knitrout}
\definecolor{shadecolor}{rgb}{0.969, 0.969, 0.969}\color{fgcolor}

{\centering \includegraphics[width=\linewidth]{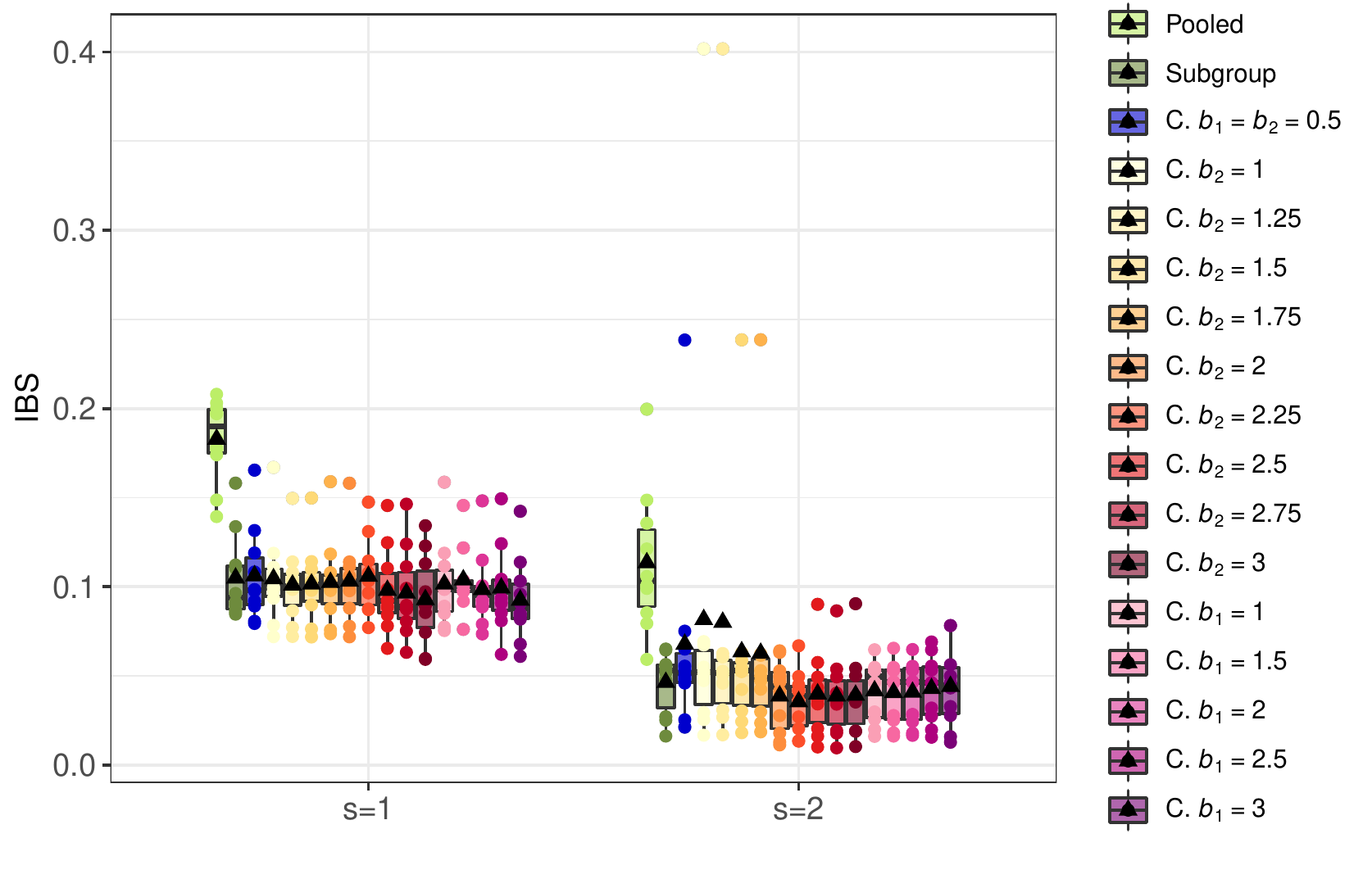} 

}

\end{knitrout}
\caption{Integrated Brier Scores (IBS) across all ten test sets for both subroups (based on the Median Probability Model). CoxBVS-SL is abbreviated by "C.". The black triangle within each boxplot represents the mean value.} 
\label{fig:GBibsMPM}
\end{figure}

When we compare all models with regard to prediction accuracy in Figure~\ref{fig:GBibsMPM} and Supplementary Figure~\ref{fig:GBibsBMA}, we again see competitive performance for the Subgroup and CoxBVS-SL model whereas Pooled is clearly worse.
We can observe a tendency towards slightly improved prediction accuracy for increasing values of $b_2$. 

\begin{figure}[!htb]
\begin{knitrout}
\definecolor{shadecolor}{rgb}{0.969, 0.969, 0.969}\color{fgcolor}

{\centering \includegraphics[width=\linewidth]{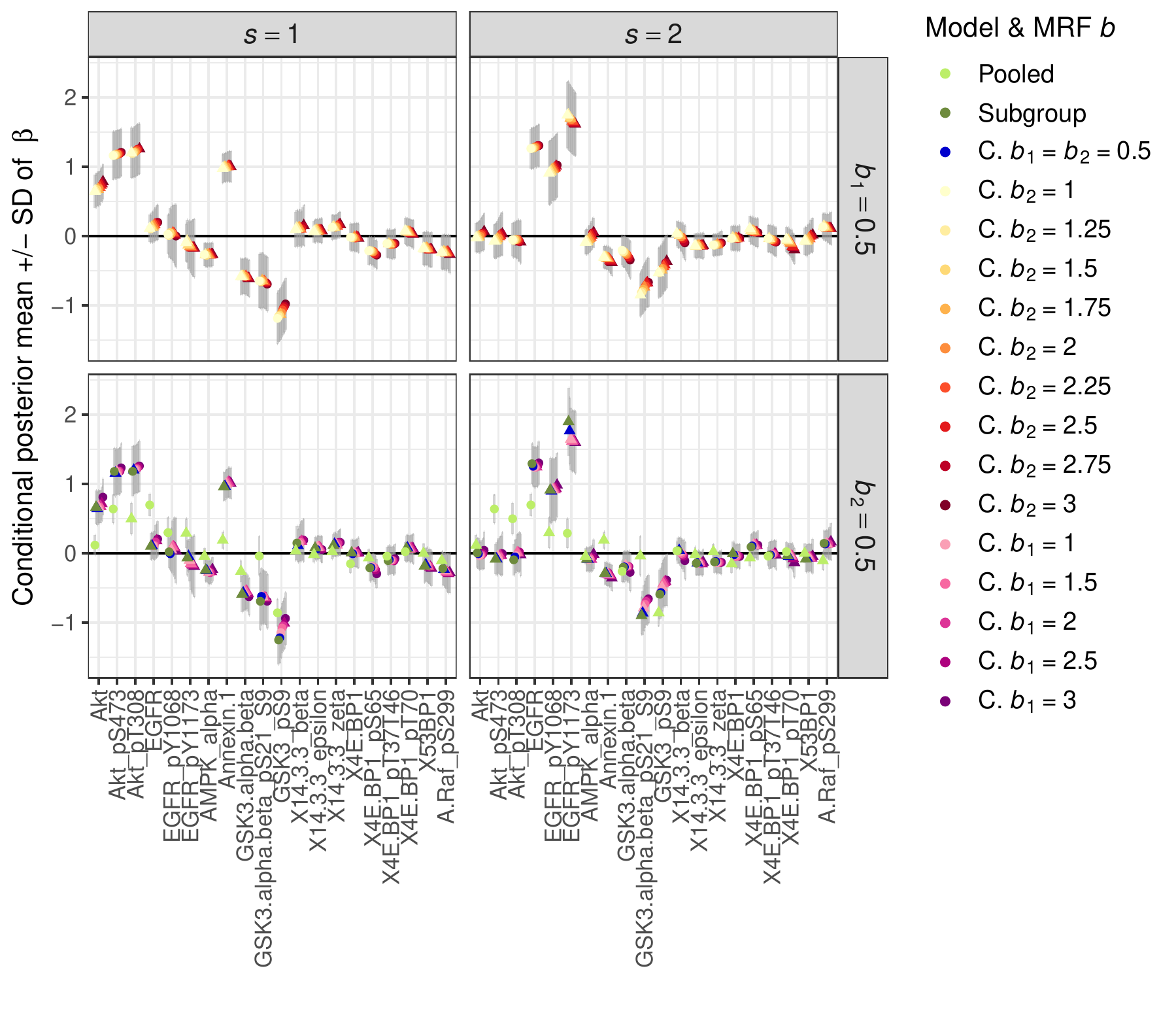} 

}

\end{knitrout}
\caption{Conditional posterior means (conditional on $\gamma = 1$) and standard deviations (SD) of the regression coefficients of all 20 proteins in both subgroups (averaged across all training sets). The different colors represent the models or parameter values of $b_1$ and $b_2$ in CoxBVS-SL (abbreviated by "C."). The plot symbol indicates whether a protein is selected (triangle) or not (circular point).} 
\label{fig:GBcondBetas}
\end{figure}

Finally, we assess the impact of increasing values of $b_2$ on the subgraph $\boldsymbol{G}_{12}$ linking both subgroups.
The corresponding marginal posterior edge selection probabilities are depicted in Supplementary Figure~\ref{fig:GBgraphG12}.
When $b_2$ becomes larger first, the posterior edge selection probabilities of proteins 8, 10 and 11 with opposite or joint effects in both subgroups increase, followed by the first six proteins with subgroup-specific effects and protein 9 with joint effect.
The posterior edge selection probabilities of the noise proteins in both subgroups remain at the prior mean and only start to increase slightly when $b_2\geq 2.5$.
Proteins 7 and 9 have much smaller posterior edge selection probabilities than the other proteins with opposite or joint effects, which fits to previous findings.

When $b_1$ becomes larger, the marginal posterior edge selection probabilities in the subgraphs $\boldsymbol{G}_{11}$ and $\boldsymbol{G}_{22}$ show no visible changes. 
In $\boldsymbol{G}_{12}$ they increase for some proteins 
however, to a much lesser extent than for larger $b_2$.

\section{Discussion} 
\label{sec:discuss}

We consider the situation of different, possibly heterogeneous patients subgroups with survival endpoint and continuous molecular measurements such as gene expression data.
When building a separate risk prediction model for each subgroup, it is important to consider heterogeneity but at the same time it can be reasonable to allow sharing information across subgroups to increase power, in particular when the sample sizes are small.
For this situation we propose a hierarchical Cox model with stochastic search variable selection prior.
To achieve higher power in variable selection and better prediction performance, we use an MRF prior instead of the standard Bernoulli prior for the latent variable selection indicators $\boldsymbol{\gamma}$.
The MRF prior leads to higher selection probabilities for genes that are related in an undirected graph. 
We use this graph to link genes across different subgroups and thereby borrow information between subgroups.
Genes that are simultaneously prognostic in different subgroups have a higher probability of being selected into the respective subgroup Cox models.
As a side aspect, the graph in the MRF prior also allows us to estimate a network between genes within each subgroup providing indications of functionally related genes and pathways. 
Here, genes that are conditionally dependent have a higher selection probability.

In the simulations and the case study we compared our proposed CoxBVS-SL model to the standard approach with independent Bernoulli prior for $\boldsymbol{\gamma}$ represented by the Subgroup and Pooled model. 
Simulations showed that the Pooled model performed worst in terms of variable selection and prediction accuracy. It averaged the effects across both subgroups and thus, led to biased estimates.
CoxBVS-SL had more power in variable selection and slightly better prediction performance than Subgroup when the sample size was small. For $n>p$ both models were competitive.

In further simulations we studied the effect of increasing values of $b_2$ representing the weight that is given to the subgraph $\boldsymbol{G}_{12}$ in the MRF prior of CoxBVS-SL and compared the results to the Sub-struct model where ${b_2=0}$.
When $b_2$ was small, CoxBVS-SL and Sub-struct performed very similarly.
Thus, the subgraph linking both subgroups had only a small influence on the results compared to the conditional dependencies among covariates within each subgroup (subgraphs $\boldsymbol{G}_{11}$ and $\boldsymbol{G}_{22}$). 
For larger values of $b_2$ prediction performance slightly improved and power in variable selection increased but on the other hand, there was a tendency towards false positive variables.

In previous simulations we increased the weight for $\boldsymbol{G}_{12}$ by choosing a larger value for the prior probability of edge inclusion $\pi$ for the corresponding edge inclusion indicators $g_{12,ii}$, $i=1,\ldots,p$.
This led to larger posterior edge selection probabilities, however, for all genes and not only the ones with joint effects. 
The variable selection results did not change remarkably. 
We could observe a small increase in power for all genes which again implied a tendency towards false positives.
We can conclude that a proper choice of $b$ (and $a$) in the MRF prior is crucial for the results of the graph and the Cox model.

We were able to demonstrate the superiority of our proposed model over the two standard approaches.
This suggests that incorporating network information into variable selection can increase power to identify the prognostic covariates and improve prediction performance.
However, in the case study the CoxBVS-SL and Subgroup model performed similarly well (Pooled was again clearly worse).
The reason for this may be that the sample sizes in both subgroups were relatively large, in particular $n>p$. 
Simulations had shown that CoxBVS-SL outperformed Subgroup only when $n \leq p$ and otherwise was competitive.

Due to computation time, we have included only up to 200 variables so far and the analysis of many thousands of genes is not (yet) feasible.
An advantage of the CoxBVS-SL model is that it does not require prior knowledge of the graph among the covariates and between subgroups. 
It accounts for uncertainty over both variable and graph selection.
In situations where pathway information is available and the graph structure is known, it is possible to incorporate this structural information in the MRF prior via a fixed graph.

\section*{Acknowledgements}

This work was supported by Deutsche Forschungsgemeinschaft (DFG) within the Collaborative Research Center SFB 876 ``Providing Information by Resource-Constrained Analysis'', project C4 (Katja Ickstadt) and project A3 (J\"org Rahnenf\"uhrer), and by the Norwegian Research Council's center for research-based innovation ``BigInsight'', project number 237718 (Manuela Zucknick).


\section*{Supplementary Materials}

Additional supporting information referenced in sections~2,~3 and~4 are available with this paper.
R source code for the models described in this paper and the preprocessed Glioblastoma data are available on GitHub: \\
{\small \url{https://github.com/KatrinMadjar/CoxBVS-SL.git}}.


\bibliographystyle{unsrt}   



\newpage
\appendix

\section*{\centering Supplementary Materials}

\section*{Details of the MCMC algorithm}

In the following, steps 1 to 4 of the MCMC sampling scheme in section \ref{sec:MethMCMC} are explained in more detail.

\subsection*{Step 1: Update of $\boldsymbol{\Omega}_{ss}$}

The block Gibbs sampler proposed by \cite{wang_scaling_2015} is used to update $\boldsymbol{\Omega}_{ss}$ for subgroups $s=1,...,S$.
The conditional distribution of $\boldsymbol{\Omega}_{ss}$ is
\begin{align*}
p(\boldsymbol{\Omega}_{ss}|\boldsymbol{G}_{ss},\boldsymbol{X}_s)
&\propto p(\boldsymbol{X}_s|\boldsymbol{\Omega}_{ss}) \cdot p(\boldsymbol{\Omega}_{ss}|\boldsymbol{G}_{ss}) \\
&\propto |\boldsymbol{\Omega}_{ss}|^{n_s/2} \exp\{ -\frac{1}{2} \text{tr}(\boldsymbol{S}_s \boldsymbol{\Omega}_{ss}) \} 
\cdot  \prod_{i<j} \exp\{-\frac{1}{2} \frac{\omega_{ss,ij}^2}{\nu^2_{g_{ss,ij}}} \} 
\cdot \prod_{i} \exp\{-\frac{\lambda}{2} \omega_{ss,ii} \}   \, . 
\end{align*}
Consider the following partitions
\small{
\[
\boldsymbol{\Omega}_{ss} = 
\left(
\begin{array}{c:c}
  \widetilde{\boldsymbol{\Omega}}_{11} & \widetilde{\boldsymbol{\omega}}_{12} \\
	  \hdashline
\widetilde{\boldsymbol{\omega}}'_{12} & \widetilde{\omega}_{22} \\
\end{array}
\right)
= 
\left(
\begin{array}{cccc:c}
 \omega_{ss,11} & \omega_{ss,12} & \ldots & \omega_{ss,1(p-1)} & \omega_{ss,1p}  \\
\omega_{ss,12} & \omega_{ss,22} & \ldots  & \omega_{ss,2(p-1)} & \omega_{ss,2p} \\
\vdots & \vdots & \ddots & \vdots  & \vdots  \\
\omega_{ss,1(p-1)} &  \omega_{ss,2(p-1)} & \ldots & \omega_{ss,(p-1)(p-1)} & \omega_{ss,(p-1)p} \\
\hdashline
\omega_{ss,1p} &  \omega_{ss,2p} & \ldots & \omega_{ss,(p-1)p} & \omega_{ss,pp} \\
\end{array}
\right)
\]}
and analogously
\small{
\[
\boldsymbol{S}_s = \boldsymbol{X}_s'\boldsymbol{X}_s =
\left(
\begin{array}{c:c}
  \widetilde{\boldsymbol{S}}_{11} & \widetilde{\boldsymbol{s}}_{12} \\
	  \hdashline
\widetilde{\boldsymbol{s}}'_{12} & \widetilde{s}_{22} \\
\end{array}
\right) \, , \quad
\boldsymbol{V}_s = (\nu^2_{g_{ss,ij}}) =
\left(
\begin{array}{c:c}
  \widetilde{\boldsymbol{V}}_{11} & \widetilde{\boldsymbol{v}}_{12} \\
	  \hdashline
\widetilde{\boldsymbol{v}}'_{12} & 0 \\
\end{array}
\right) \, ,
\]}
where $\boldsymbol{V}_s$ is a $(p \times p)$ symmetric matrix with zeros on the diagonal.
For the block update of $\boldsymbol{\Omega}_{ss}$ focus on the last column (and row) of $\boldsymbol{\Omega}_{ss}$:
$(\widetilde{\boldsymbol{\omega}}_{12} , \widetilde{\omega}_{22} )$
with ${\widetilde{\boldsymbol{\omega}}_{12} = (\omega_{ss,1p} ,\omega_{ss,2p} ,...,\omega_{ss,(p-1)p})'}$, 
$\widetilde{\omega}_{22} = \omega_{ss,pp}$. \\
The conditional distribution of the last column of $\boldsymbol{\Omega}_{ss}$ is
\[
p( \widetilde{\boldsymbol{\omega}}_{12} , \widetilde{\omega}_{22}| \boldsymbol{X}_s, \boldsymbol{G}_{ss}, \widetilde{\boldsymbol{\Omega}}_{11} )  \propto 
\big(\widetilde{\omega}_{22} - \widetilde{\boldsymbol{\omega}}_{12}' \widetilde{\boldsymbol{\Omega}}^{-1}_{11} \widetilde{\boldsymbol{\omega}}_{12} \big)^{n_s/2} 
\cdot \exp\Big\{ -\frac{1}{2} \left[ \widetilde{\boldsymbol{\omega}}_{12}' \text{diag}(\widetilde{\boldsymbol{v}}^{-1}_{12})\widetilde{\boldsymbol{\omega}}_{12} + 2 \widetilde{\boldsymbol{s}}'_{12} \widetilde{\boldsymbol{\omega}}_{12} + (\widetilde{s}_{22}+\lambda) \widetilde{\omega}_{22} \right] \Big\} \, . 
\]
Consider the following transformations
\[
\boldsymbol{u} = \widetilde{\boldsymbol{\omega}}_{12} \, , \quad 
v = \widetilde{\omega}_{22} - \widetilde{\boldsymbol{\omega}}_{12}' \widetilde{\boldsymbol{\Omega}}^{-1}_{11} \widetilde{\boldsymbol{\omega}}_{12}  \, .
\] 
Then the conditional distribution is
\[
p(\boldsymbol{u},v| \boldsymbol{X}_s, \boldsymbol{G}_{ss}, \widetilde{\boldsymbol{\Omega}}_{11} )  \propto 
\underbrace{v^{n_s/2} \exp\Big\{ -\frac{\widetilde{s}_{22}+\lambda}{2} v \Big\}}_{ (*_1) }   
\cdot \underbrace{\exp\Big\{ -\frac{1}{2} \Big[ \boldsymbol{u}' \underbrace{\left(\text{diag}(\widetilde{\boldsymbol{v}}^{-1}_{12})+(\widetilde{s}_{22}+\lambda) \widetilde{\boldsymbol{\Omega}}^{-1}_{11} \right)}_{=\boldsymbol{C}^{-1}} \boldsymbol{u} + 2 \widetilde{\boldsymbol{s}}'_{12}\boldsymbol{u} \Big] \Big\}}_{(*_2)}   
\]
$(*_1) \; \propto \mathcal{G}(v | \frac{n_s}{2}+1, \frac{\widetilde{s}_{22} + \lambda}{2})$, \\
$(*_2) \; \propto \mathcal{N}(\boldsymbol{u} | -\boldsymbol{C}\widetilde{\boldsymbol{s}}_{12}, \boldsymbol{C})$. 
\\
Permuting any column in $\boldsymbol{\Omega}_{ss}$ to be updated to the last one leads to a block Gibbs sampler for the update of $\boldsymbol{\Omega}_{ss}$.


\subsection*{Step 2: Update of $\boldsymbol{G}$}

Update all elements in $\boldsymbol{G}$ iteratively with Gibbs sampler from their conditional distributions.
All elements $g_{rs,ij}$ are assumed independent Bernoulli a priori with ${p(g_{rs,ij}=1)=\pi}$ and ${p(g_{rs,ij}=0)=1-\pi}$. 

Update $g_{rs,ii}$, ${r,s=1,...,S}$, $r<s$, $i=1,...,p$ (edges between the same gene in different subgroups) from the conditional distribution
\begin{align}
p(g_{rs,ii} | \boldsymbol{G}_{-rs,ii}, \boldsymbol{\gamma}) &= \frac{p(g_{rs,ii}) \cdot p(\boldsymbol{\gamma}| \boldsymbol{G}_{-rs,ii}, g_{rs,ii})}{\sum_{g_{rs,ii} \in \{0,1\}} p(g_{rs,ii}) \cdot p(\boldsymbol{\gamma}| \boldsymbol{G}_{-rs,ii}, g_{rs,ii})} \, , \notag 
\end{align}
where $\boldsymbol{G}_{-rs,ii}$ denotes all elements in $\boldsymbol{G}$ except for $g_{rs,ii}$.
Accept $g_{rs,ii} =1$ with probability
\[
p(g_{rs,ii} =1| \boldsymbol{G}_{-rs,ii}, \boldsymbol{\gamma} ) = \frac{w_a}{w_a + w_b}  , 
\]
where
\vspace{-0.5cm}
\begin{align}
w_a &= \pi \cdot \exp(a \boldsymbol{1}_{pS}' \boldsymbol{\gamma}  + b \boldsymbol{\gamma}'\boldsymbol{G} \boldsymbol{\gamma})|_{g_{rs,ii}=1} \notag \\
w_b &= (1-\pi) \cdot \exp(a \boldsymbol{1}_{pS}' \boldsymbol{\gamma}  + b \boldsymbol{\gamma}'\boldsymbol{G} \boldsymbol{\gamma})|_{g_{rs,ii}=0}  \; . \notag
\end{align}
This means, update $g_{rs,ii}$ as follows: \; 
$g_{rs,ii}=\begin{cases}
  1,  & \text{if} \; u < \frac{w_a}{w_a + w_b} , \;  u \sim \mathcal{U}[0,1] \\
  0, & \text{else} \, .
\end{cases}$ 
\vspace{0.5cm}\\
Update $g_{ss,ij}$, ${s=1,...,S}$, ${i,j=1,...,p}$, ${i<j}$ (edges between different genes in the same subgroup) from the conditional distribution 
\begin{align*}
p(g_{ss,ij} | \boldsymbol{G}_{-ss,ij}, \omega_{ss,ij}, \boldsymbol{\gamma}) 
&= \frac{p(g_{ss,ij}) \cdot p(\omega_{ss,ij}, \boldsymbol{\gamma}| \boldsymbol{G}_{-ss,ij}, g_{ss,ij})}{\sum_{g_{ss,ij} \in \{0,1\}} p(g_{ss,ij}) \cdot p(\omega_{ss,ij}, \boldsymbol{\gamma}| \boldsymbol{G}_{-ss,ij}, g_{ss,ij})} \\
& \propto p(g_{ss,ij}) \cdot p(\omega_{ss,ij}| g_{ss,ij}) \cdot p( \boldsymbol{\gamma}| \boldsymbol{G}_{-ss,ij}, g_{ss,ij})
\, ,
\end{align*}
where $\boldsymbol{G}_{-ss,ij}$ denotes all elements in $\boldsymbol{G}$ except for $g_{-ss,ij}$.
Accept $g_{ss,ij} =1$ with probability
\[
p(g_{ss,ij} =1| \boldsymbol{G}_{-ss,ij}, \omega_{ss,ij}, \boldsymbol{\gamma} ) = \frac{w_a}{w_a + w_b} , 
\]
where
\vspace{-0.5cm}
\begin{align}
w_a &= \pi \cdot \mathcal{N}(\omega_{ss,ij}| 0, \nu^2_1) \cdot \exp(a \boldsymbol{1}_{pS}' \boldsymbol{\gamma}  + b \boldsymbol{\gamma}'G \boldsymbol{\gamma})|_{g_{ss,ij}=1}  \notag \\
w_b &= (1-\pi) \cdot \mathcal{N}(\omega_{ss,ij}| 0, \nu^2_0) \cdot \exp(a \boldsymbol{1}_{pS}' \boldsymbol{\gamma}  + b \boldsymbol{\gamma}'G \boldsymbol{\gamma})|_{g_{ss,ij}=0} . \notag
\end{align}

\subsection*{Step 3: Update of $\boldsymbol{\gamma}$}

Update $\gamma_{s,i}$, $s=1,...,S$, $i=1,...,p$, with Gibbs sampler from the conditional distribution
\begin{align}
p(\gamma_{s,i} | \boldsymbol{\gamma}_{-s,i}, \boldsymbol{G}, \beta_{s,i}) &= \frac{p(\gamma_{s,i}, \beta_{s,i}|\boldsymbol{\gamma}_{-s,i}, \boldsymbol{G})}{\sum_{\gamma_{s,i}\in\{0,1\}} p(\gamma_{s,i}, \beta_{s,i}|\boldsymbol{\gamma}_{-s,i}, \boldsymbol{G})} \notag \\
&= \frac{p(\gamma_{s,i}|\boldsymbol{\gamma}_{-s,i}, \boldsymbol{G}) \cdot p(\beta_{s,i}|\gamma_{s,i},\boldsymbol{\gamma}_{-s,i}, \boldsymbol{G})}{\sum_{\gamma_{s,i}\in\{0,1\}} p(\gamma_{s,i}|\boldsymbol{\gamma}_{-s,i}, \boldsymbol{G}) \cdot p(\beta_{s,i}|\gamma_{s,i},\boldsymbol{\gamma}_{-s,i}, \boldsymbol{G})} \notag \\
 & = 
\frac{p(\gamma_{s,i},\boldsymbol{\gamma}_{-s,i} | \boldsymbol{G}) \cdot p(\beta_{s,i}|\gamma_{s,i})}{\sum_{\gamma_{s,i}\in\{0,1\}} p(\gamma_{s,i},\boldsymbol{\gamma}_{-s,i} | \boldsymbol{G}) \cdot p(\beta_{s,i}|\gamma_{s,i})} \, , \notag 
\end{align}
where $\boldsymbol{\gamma}_{-s,i}$ denotes all elements in $\boldsymbol{\gamma}$ except for $\gamma_{s,i}$.
Accept $\gamma_{s,i}=1$ with probability
\[
p(\gamma_{s,i} =1 | \boldsymbol{\gamma}_{-s,i}, \boldsymbol{G}, \beta_{s,i}) = \frac{w_a}{w_a + w_b} , 
\]
where
\vspace{-0.5cm}
\begin{align}
w_a &= 
  \exp(a \boldsymbol{1}_{pS}' \boldsymbol{\gamma}  + b \boldsymbol{\gamma}'\boldsymbol{G} \boldsymbol{\gamma})|_{\gamma_{s,i}=1} \cdot \mathcal{N}(\beta_{s,i}|0,c^2 \tau^2)
 \notag \\
w_b &= 
 \exp(a \boldsymbol{1}_{pS}' \boldsymbol{\gamma}  + b \boldsymbol{\gamma}'\boldsymbol{G} \boldsymbol{\gamma})|_{\gamma_{s,i}=0} \cdot \mathcal{N}(\beta_{s,i}|0, \tau^2)  .
\notag
\end{align}

\subsection*{Step 4: Update of $\boldsymbol{\beta}$}

A random walk Metropolis-Hastings algorithm with adaptive jumping rule as proposed by \cite{lee_bayesian_2011} is used to update $\beta_{s,i}$ for $s=1,...,S$ and $i=1,...,p$.
The full conditional posterior distribution of $\beta_{s,i}$ is
\begin{eqnarray*}
\lefteqn{p(\beta_{s,i}|\boldsymbol{\beta}_{s,-i}, \boldsymbol{\gamma}_s, \boldsymbol{h}_s, \mathfrak{D}_s)} \\
& \propto & L(\mathfrak{D}_s|\boldsymbol{\beta}_s, \boldsymbol{h}_s) \cdot p(\boldsymbol{\beta}_s|\boldsymbol{\gamma}_s)  \\
& \propto & 
\prod_{g=1}^{J_s} \left[ \exp\Big( -h_{s,g} \mspace{-20mu} \sum_{k \in \mathcal{R}_{s,g}-\mathcal{D}_{s,g}} \mspace{-20mu} \exp(\boldsymbol{\beta}_s'\boldsymbol{x}_{s,k}) \Big) \prod_{l \in \mathcal{D}_{s,g}} \Big[ 1-\exp\big( -h_{s,g} \exp(\boldsymbol{\beta}_s' \boldsymbol{x}_{s,l} ) \big) \Big] \right] 
\cdot \exp \Big( -\frac{1}{2} \boldsymbol{\beta}_s' \Sigma_{\beta_s}^{-1} \boldsymbol{\beta}_s \Big) , 
\end{eqnarray*}
where $\boldsymbol{\beta}_{s,-i}$ denotes the vector $\boldsymbol{\beta}_s$ without the $i$-th element. 
${\Sigma_{\beta_s} = \text{diag}(\sigma^2_{\beta_{s,1}},...,\sigma^2_{\beta_{s,p}})}$ with $\sigma^2_{\beta_{s,i}} = (1-\gamma_{s,i}) \cdot \tau^2 + \gamma_{s,i} \cdot c^2 \tau^2$. 

In MCMC iteration $t$ update $\beta_{s,i}$ as follows:
\begin{enumerate}
\item[(i)]
Sample a proposal $\beta_{s,i}^{(prop)}$ from a proposal distribution \,$q(\beta_{s,i}^{(prop)}|\beta_{s,i}^{(t-1)})=\mathcal{N}(\beta_{s,i}^{(prop)}|\mu_{\beta_{s,i}}^{(t-1)}, \nu_{\beta_{s,i}}^{(t-1)})$
\item[(ii)]
Calculate the ratio of ratios
\[
r_{s,i} = \frac{p(\beta_{s,i}^{(prop)}|\boldsymbol{\beta}_{s,-i}^{(t-1)}, \boldsymbol{\gamma}_s^{(t-1)}, \boldsymbol{h}_s^{(t-1)}, \mathfrak{D}_s)/ q(\beta_{s,i}^{(prop)}|\beta_{s,i}^{(t-1)}) }{p(\beta_{s,i}^{(t-1)}|\boldsymbol{\beta}_{s,-i}^{(t-1)}, \boldsymbol{\gamma}_s^{(t-1)}, \boldsymbol{h}_s^{(t-1)}, \mathfrak{D}_s)/ q(\beta_{s,i}^{(t-1)}|\beta_{s,i}^{(prop)})}
\]
\item[(iii)]
Accept the proposal $\beta_{s,i}^{(prop)}$ if $\text{min}\{r_{s,i},1\}>u$ with $u \sim \mathcal{U}[0,1]$.
\end{enumerate}
The mean 
and variance 
of the proposal distribution can be approximated based on the first and second derivative of the log conditional posterior distribution with respect to $\beta_{s,i}^{(t-1)}$.

\newpage

\section*{Supplementary Figures}

\begin{figure}[!htb] 
\begin{knitrout}
\definecolor{shadecolor}{rgb}{0.969, 0.969, 0.969}\color{fgcolor}

{\centering \includegraphics[width=\linewidth]{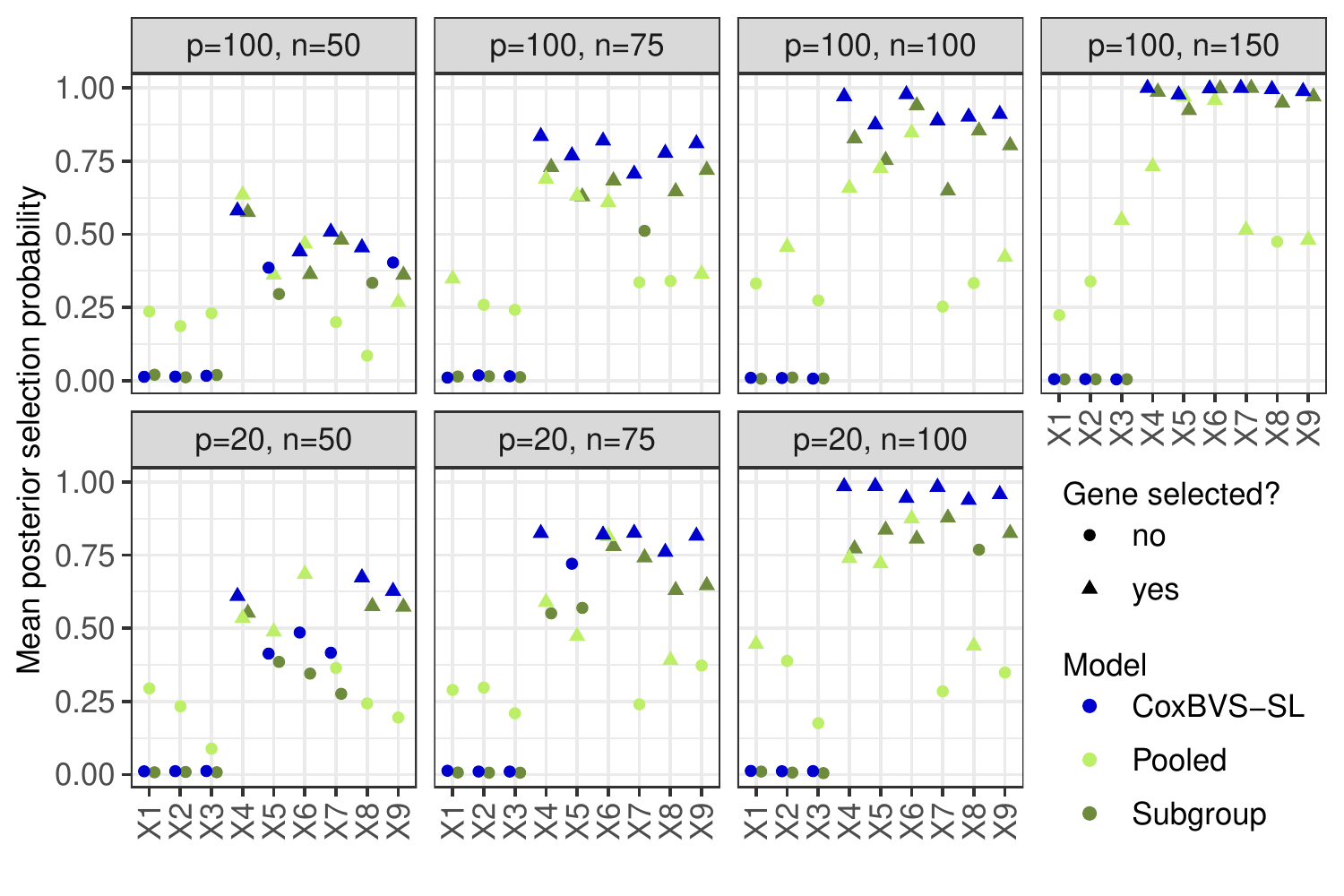} 

}

\end{knitrout}
\caption{Mean posterior selection probabilities of the first nine genes in subgroup~2 (averaged across all training sets).} 
\label{fig:BayesSim1PPI2}
\end{figure}
  
\begin{figure}[!htb] 
\begin{knitrout}
\definecolor{shadecolor}{rgb}{0.969, 0.969, 0.969}\color{fgcolor}

{\centering \includegraphics[width=\linewidth]{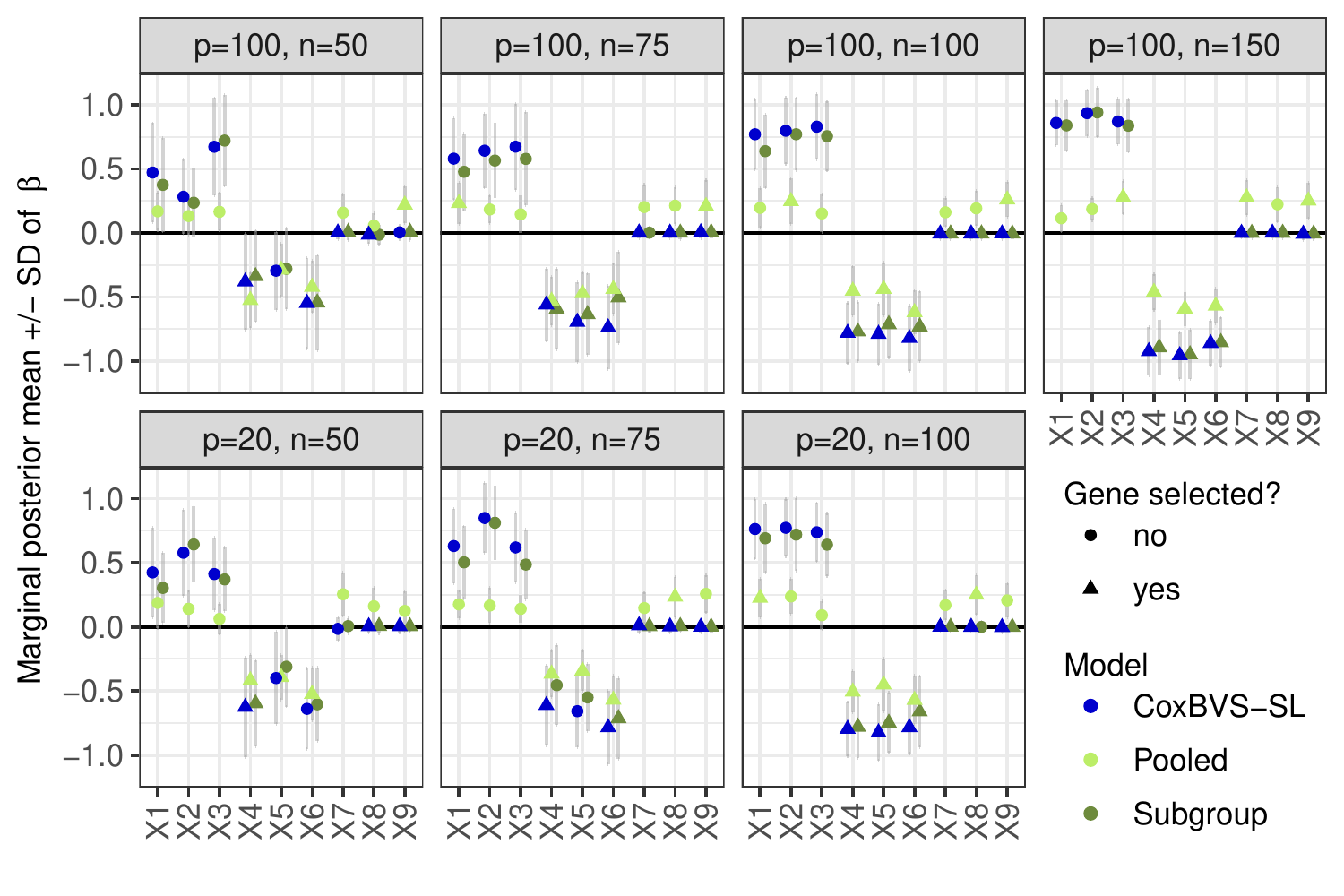} 

}

\end{knitrout}
\caption{Marginal posterior means (independent of $\gamma$) and standard deviations (SD)  of the regression coefficients of the first nine genes in subgroup~1 (averaged across all training sets). 
} 
\label{fig:BayesSim1Betas1}
\end{figure}

\begin{figure}[!htb] 
\begin{knitrout}
\definecolor{shadecolor}{rgb}{0.969, 0.969, 0.969}\color{fgcolor}

{\centering \includegraphics[width=\linewidth]{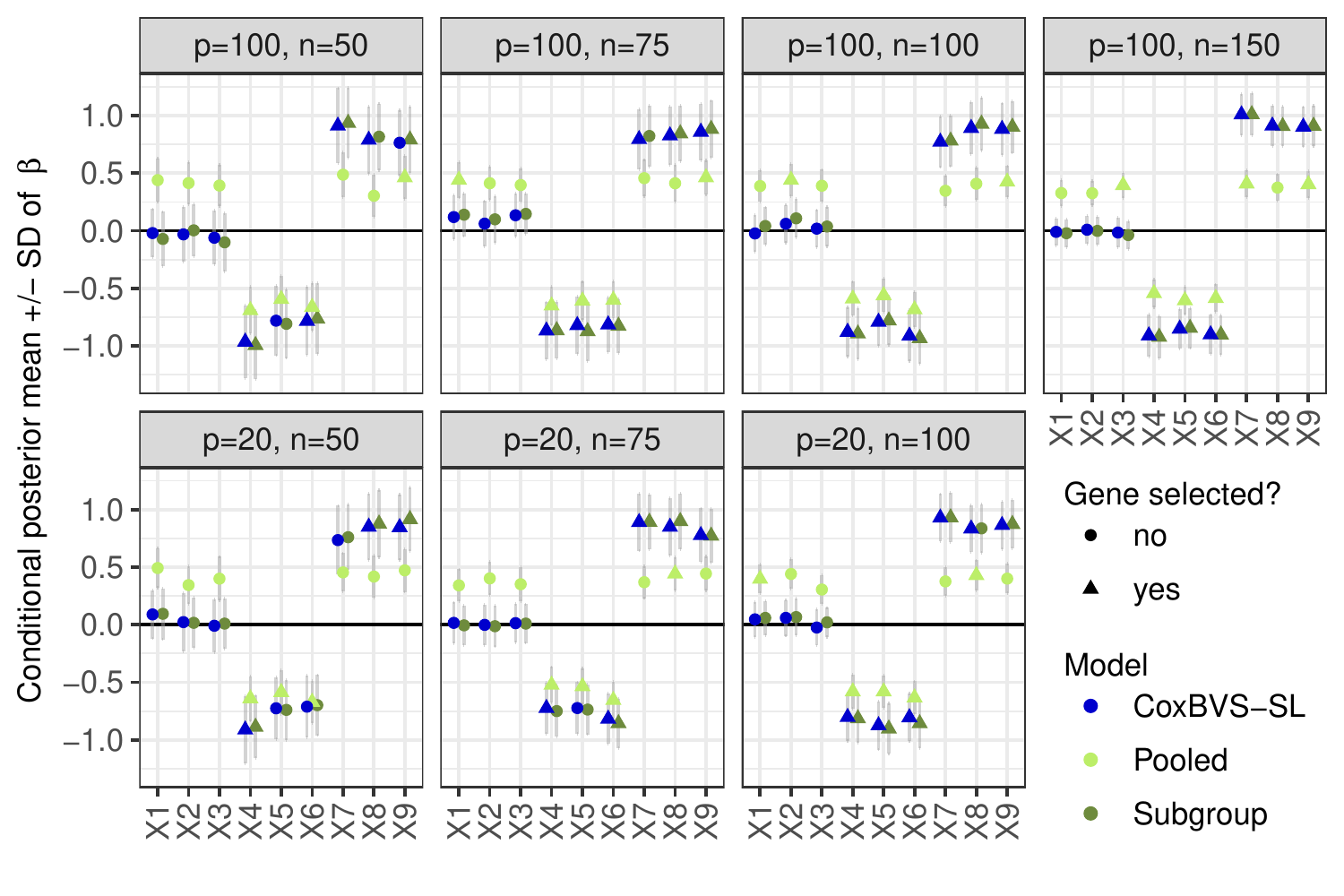} 

}

\end{knitrout}
\caption{Conditional posterior means (conditional on $\gamma=1$) and standard deviations (SD)  of the regression coefficients of the first nine genes in subgroup~2 (averaged across all training sets).} 
\label{fig:BayesSim1Betas2s2}
\end{figure}

\begin{figure}[!htb] 
\begin{knitrout}
\definecolor{shadecolor}{rgb}{0.969, 0.969, 0.969}\color{fgcolor}

{\centering \includegraphics[width=\linewidth]{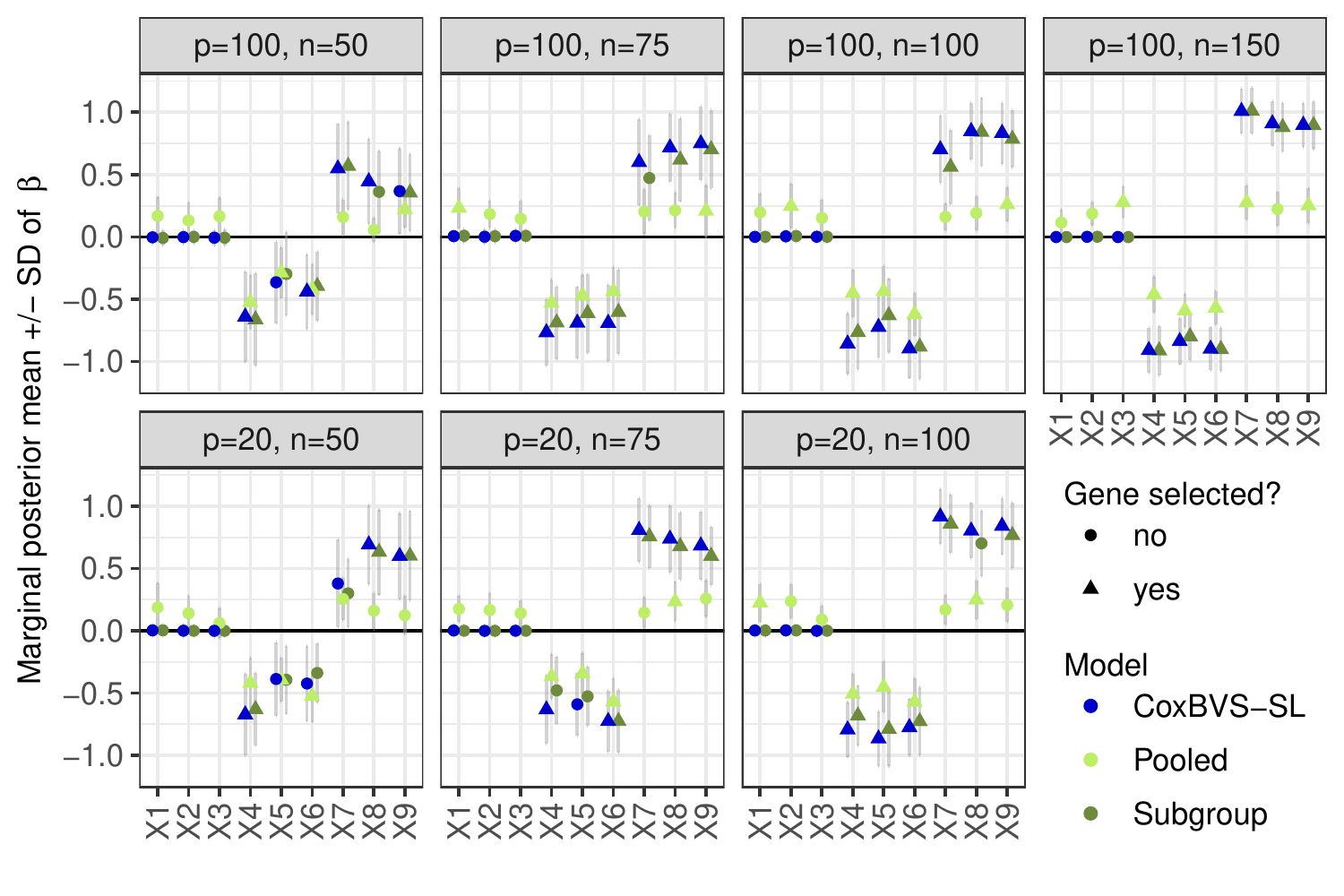} 

}

\end{knitrout}
\caption{Marginal posterior means (independent of $\gamma$) and standard deviations (SD)  of the regression coefficients of the first nine genes in subgroup~2 (averaged across all training sets).} 
\label{fig:BayesSim1Betas1s2}
\end{figure}

\begin{figure}[!htb] 
\begin{knitrout}
\definecolor{shadecolor}{rgb}{0.969, 0.969, 0.969}\color{fgcolor}

{\centering \includegraphics[width=\linewidth]{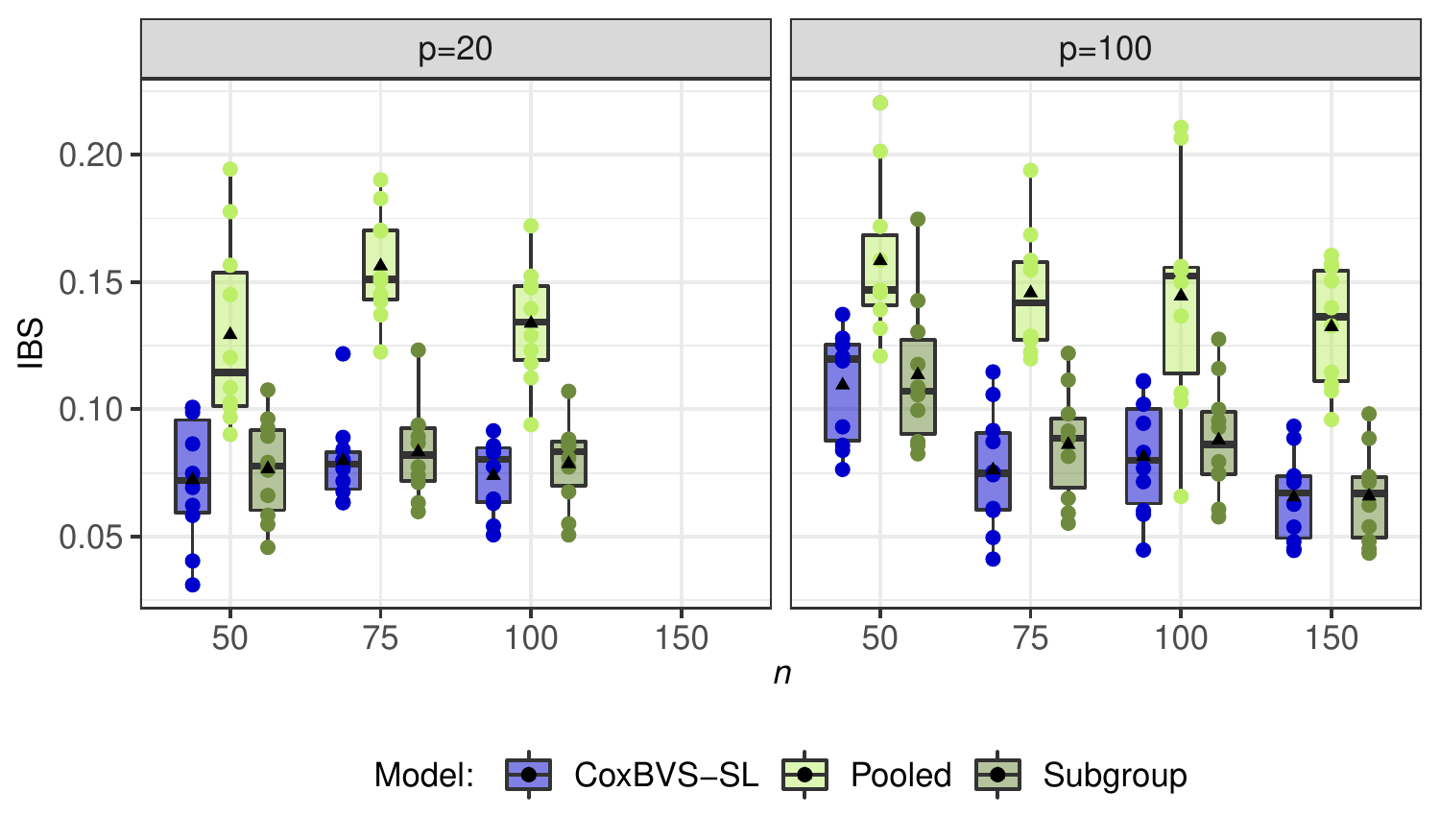} 

}

\end{knitrout}
\caption{Integrated Brier Scores (IBS) across all ten test sets for subroup~1 (IBS based on the Bayesian Model Averaging). The black triangle within each boxplot represents the mean value.} 
\label{fig:BayesSim1IBSBMA}
\end{figure}

\begin{figure}[!htb] 
\begin{knitrout}
\definecolor{shadecolor}{rgb}{0.969, 0.969, 0.969}\color{fgcolor}

{\centering \includegraphics[width=\linewidth]{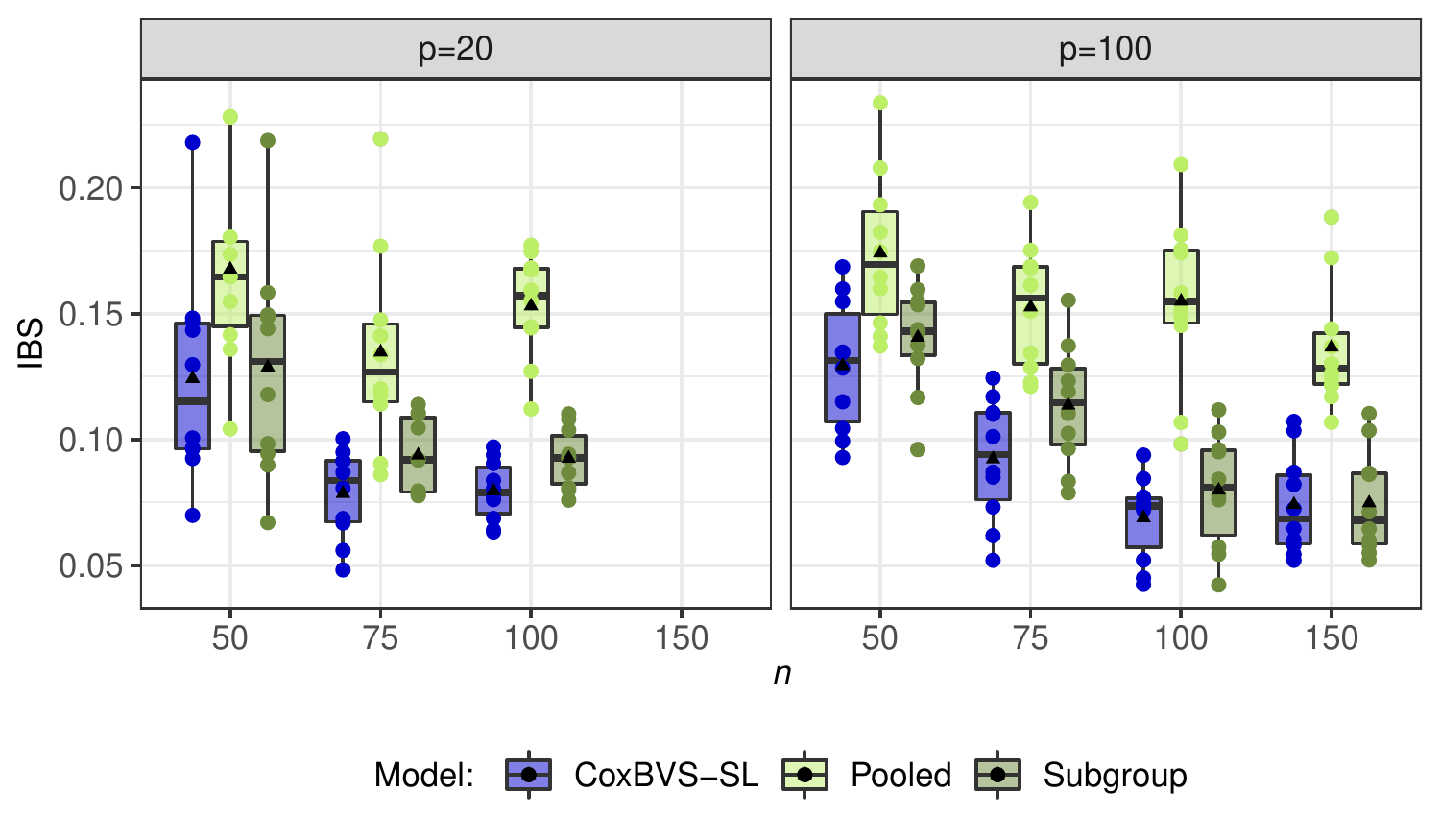} 

}

\end{knitrout}
\caption{Integrated Brier Scores (IBS) across all ten test sets for subroup~2 (IBS based on the Median Probability Model). The black triangle within each boxplot represents the mean value.} 
\label{fig:BayesSim1IBSMPMs2}
\end{figure}

\begin{figure}[!htb] 
\begin{knitrout}
\definecolor{shadecolor}{rgb}{0.969, 0.969, 0.969}\color{fgcolor}

{\centering \includegraphics[width=\linewidth]{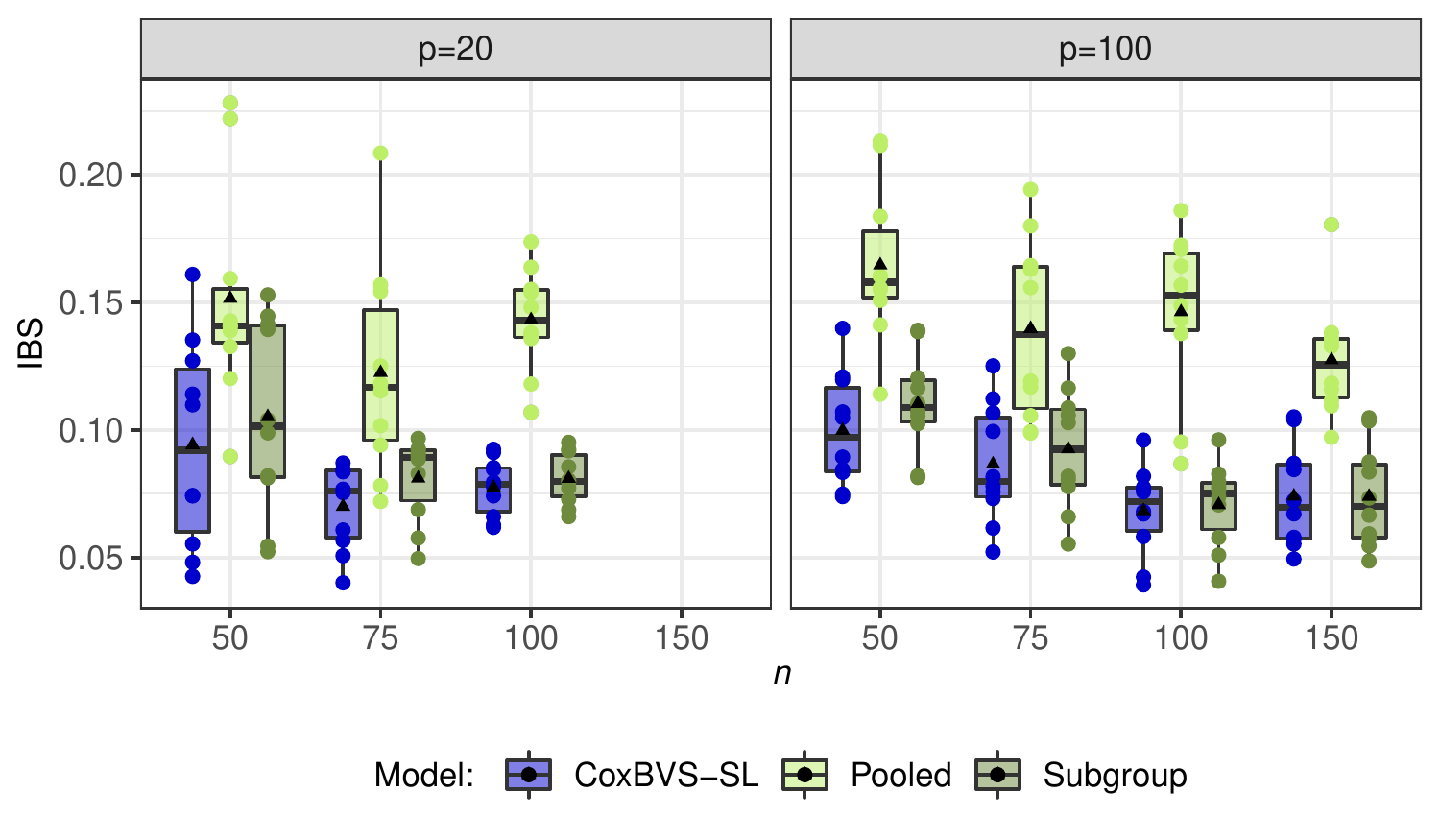} 

}

\end{knitrout}
\caption{Integrated Brier Scores (IBS) across all ten test sets for subroup~2 (IBS based on the Bayesian Model Averaging). The black triangle within each boxplot represents the mean value.} 
\label{fig:BayesSim1IBSBMAs2}
\end{figure}

\begin{figure}[!htb] 
\begin{knitrout}
\definecolor{shadecolor}{rgb}{0.969, 0.969, 0.969}\color{fgcolor}

{\centering \includegraphics[width=\linewidth]{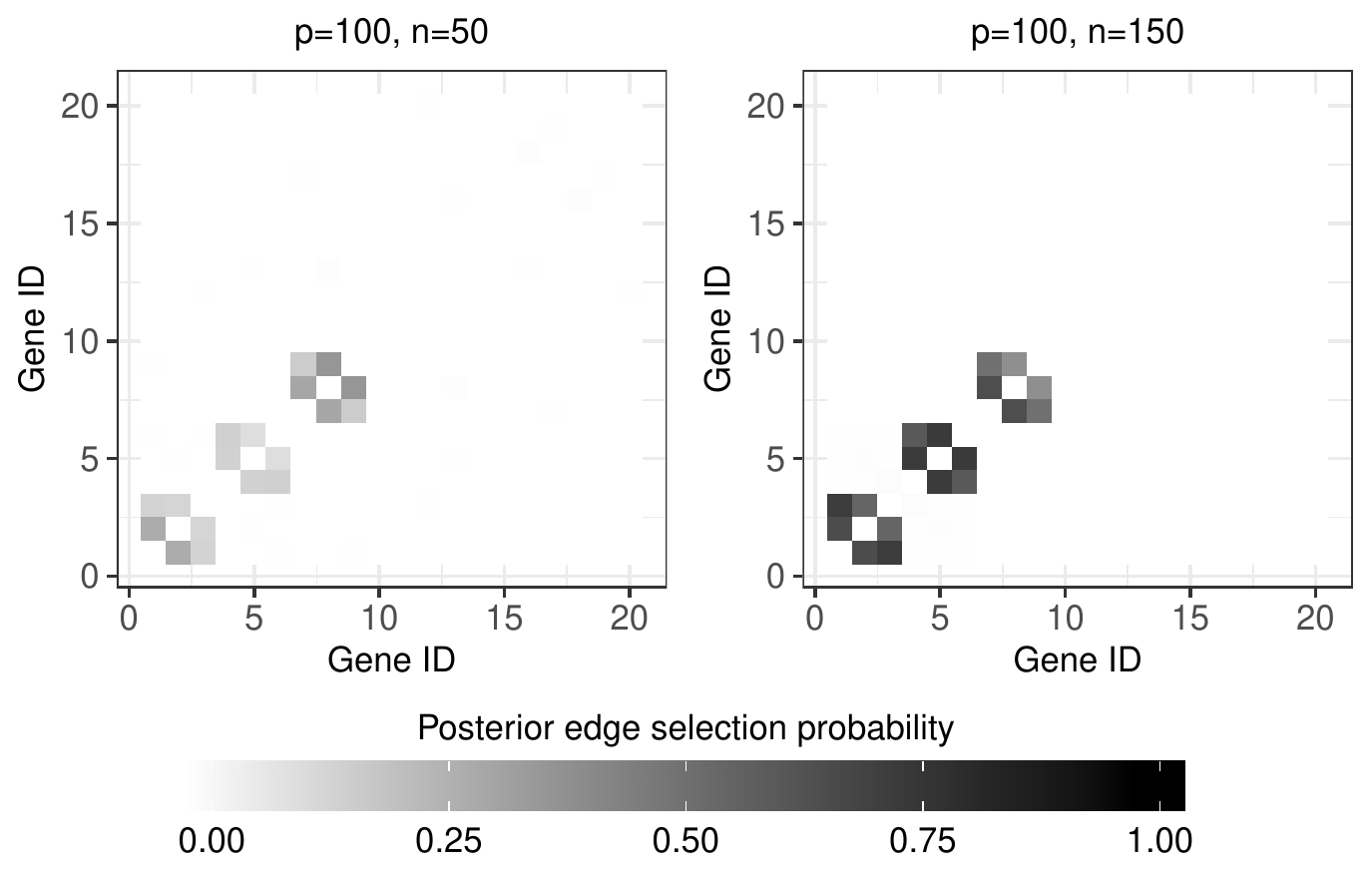} 

}

\end{knitrout}
\caption{Marginal posterior edge selection probabilities of the first 20 genes in $\boldsymbol{G}_{11}$ (averaged across all training sets) for small and large $n$. Results for subgroup~2 are very similar.}
\label{fig:BayesSim1Gss}
\end{figure}

\begin{figure}[!htb] 
\begin{knitrout}
\definecolor{shadecolor}{rgb}{0.969, 0.969, 0.969}\color{fgcolor}

{\centering \includegraphics[width=\linewidth]{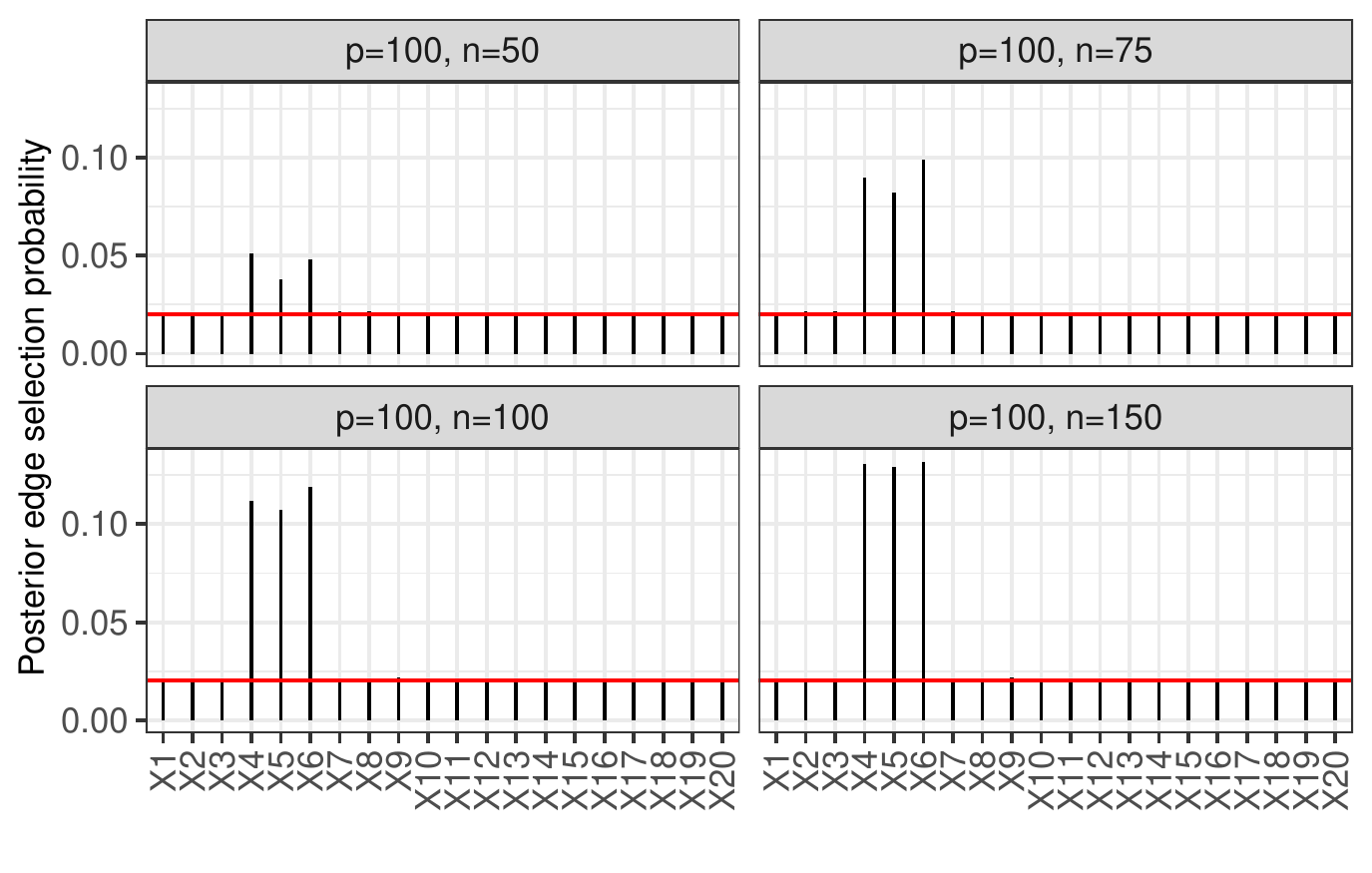} 

}

\end{knitrout}
  \caption{Marginal posterior edge selection probabilities of the first 20 genes in $\boldsymbol{G}_{12}$ (averaged across all training sets). The red line indicates the prior mean ($\pi=2/(p-1) \approx 0.02$ for $p=100$).} 
  \label{fig:BayesSim1G12}
\end{figure}



\begin{figure}[!htb] 
\begin{knitrout}
\definecolor{shadecolor}{rgb}{0.969, 0.969, 0.969}\color{fgcolor}

{\centering \includegraphics[width=\linewidth]{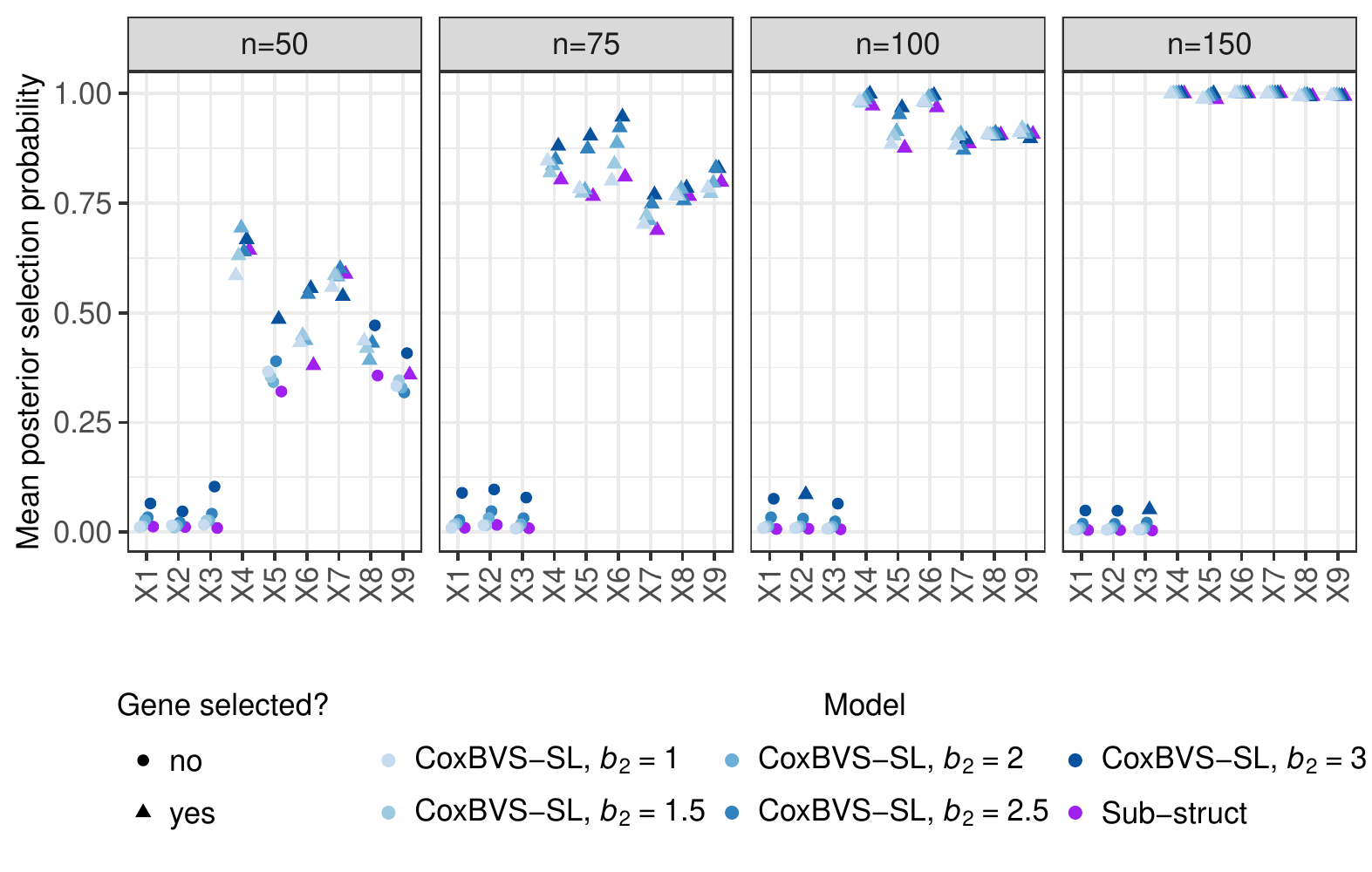} 

}

\end{knitrout}
  \caption{Mean posterior selection probabilities of the first nine genes in subgroup 2 (averaged across all training sets).} 
\label{fig:BayesSim2PPIs2}
\end{figure}

\begin{figure}[!htb] 
\begin{knitrout}
\definecolor{shadecolor}{rgb}{0.969, 0.969, 0.969}\color{fgcolor}

{\centering \includegraphics[width=\linewidth]{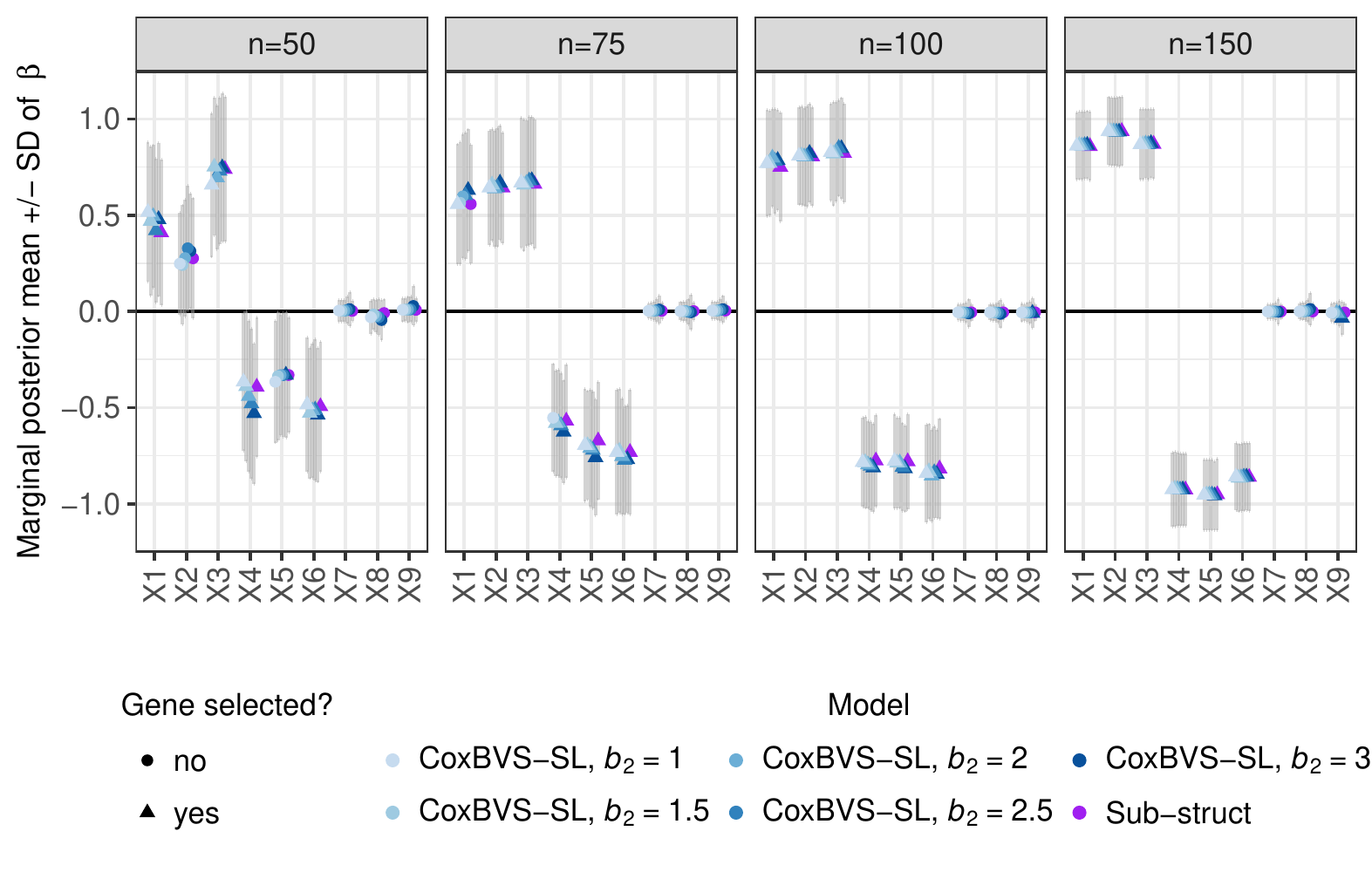} 

}

\end{knitrout}
\caption{Marginal posterior means (independent of $\gamma$) and standard deviations (SD)  of the regression coefficients of the first nine genes in subgroup~1 (averaged across all training sets). 
} 
\label{fig:BayesSim2Betas1}
\end{figure}

\begin{figure}[!htb] 
\begin{knitrout}
\definecolor{shadecolor}{rgb}{0.969, 0.969, 0.969}\color{fgcolor}

{\centering \includegraphics[width=\linewidth]{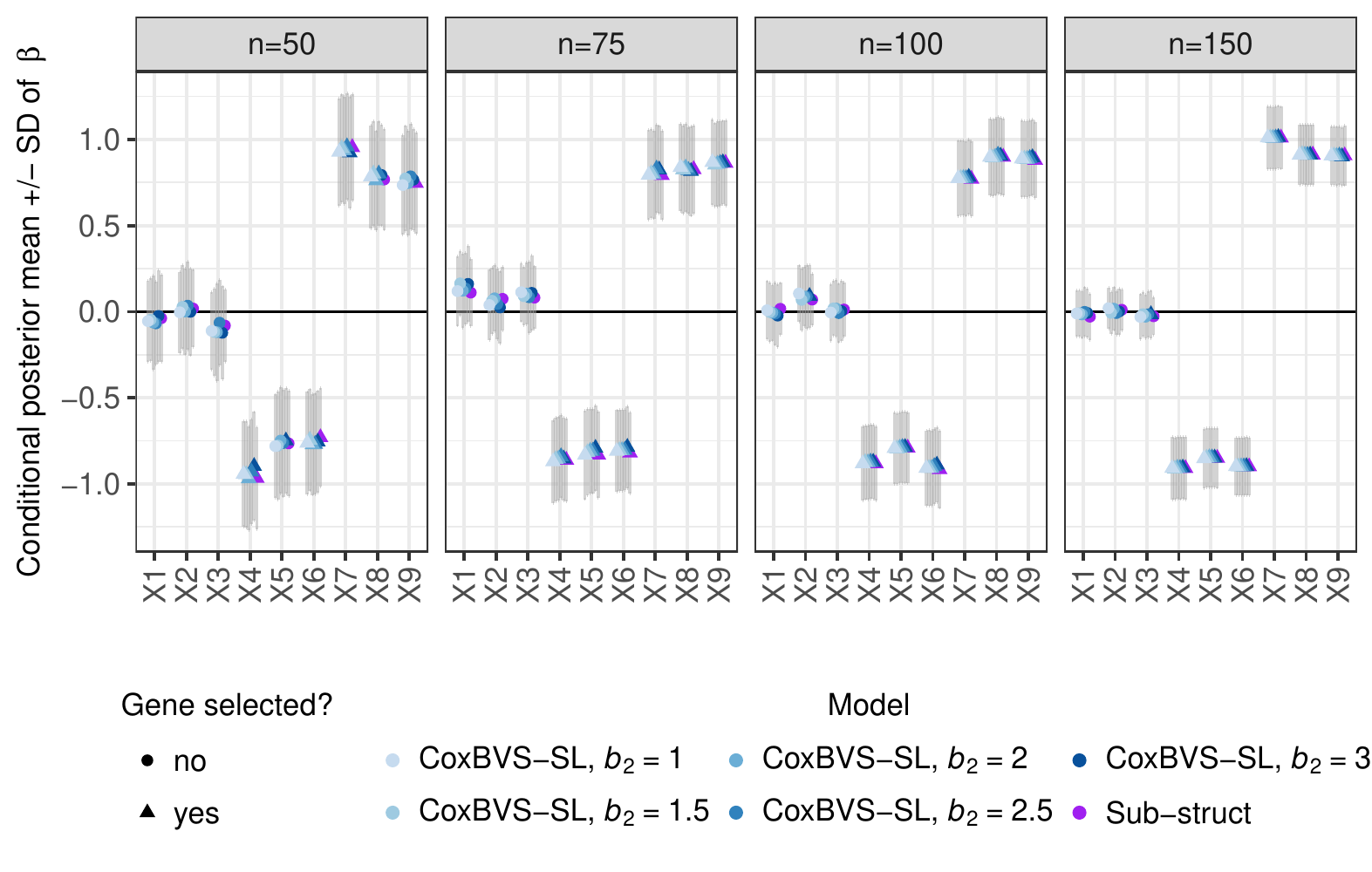} 

}

\end{knitrout}
\caption{Conditional posterior means (conditional on $\gamma=1$) and standard deviations (SD)  of the regression coefficients of the first nine genes in subgroup~2 (averaged across all training sets).} 
\label{fig:BayesSim2Betas2s2}
\end{figure}

\begin{figure}[!htb] 
\begin{knitrout}
\definecolor{shadecolor}{rgb}{0.969, 0.969, 0.969}\color{fgcolor}

{\centering \includegraphics[width=\linewidth]{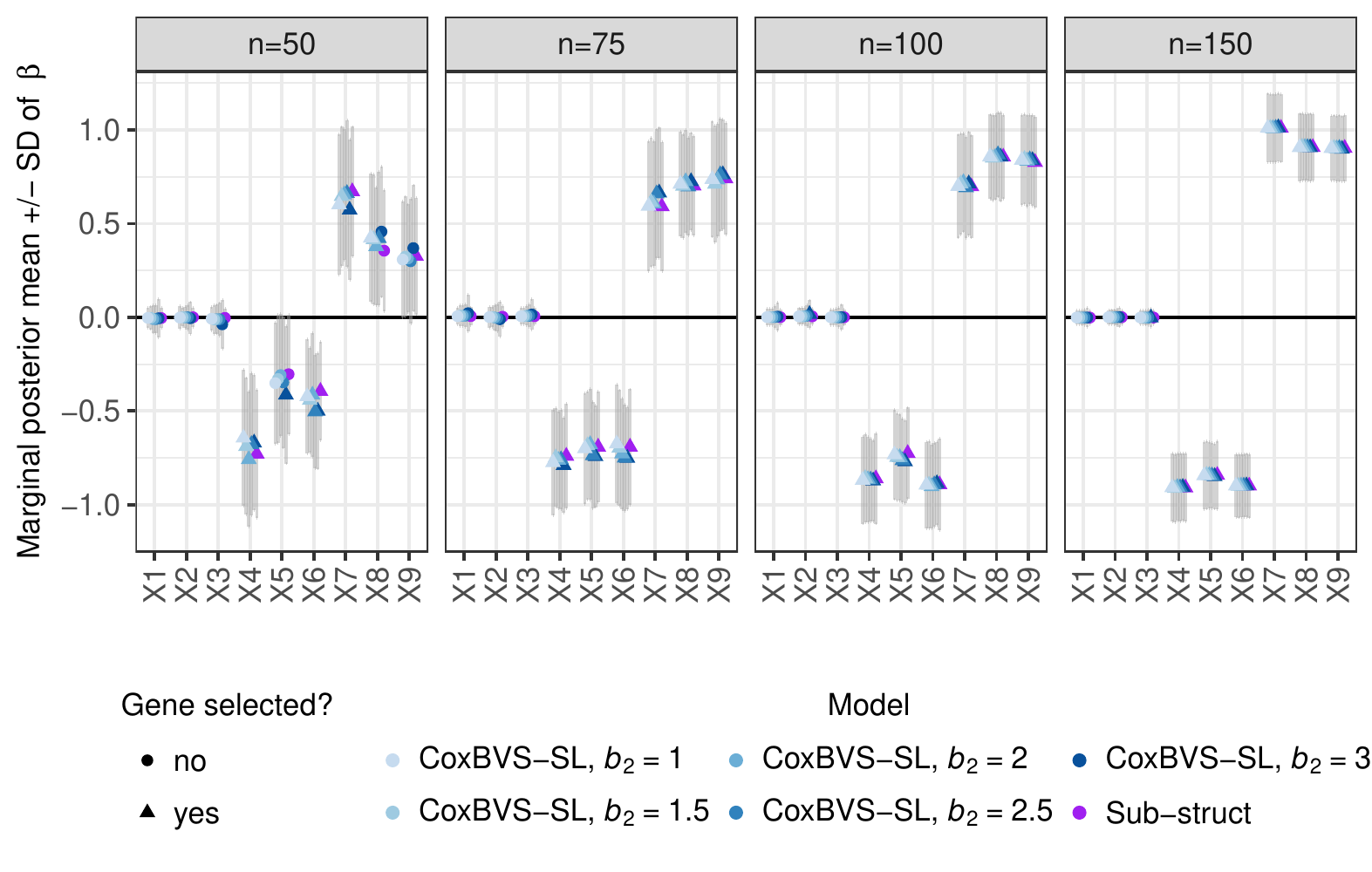} 

}

\end{knitrout}
\caption{Marginal posterior means (independent of $\gamma$) and standard deviations (SD)  of the regression coefficients of the first nine genes in subgroup~2 (averaged across all training sets).} 
\label{fig:BayesSim2Betas1s2}
\end{figure}

\begin{figure}[!htb]
\begin{knitrout}
\definecolor{shadecolor}{rgb}{0.969, 0.969, 0.969}\color{fgcolor}

{\centering \includegraphics[width=\linewidth]{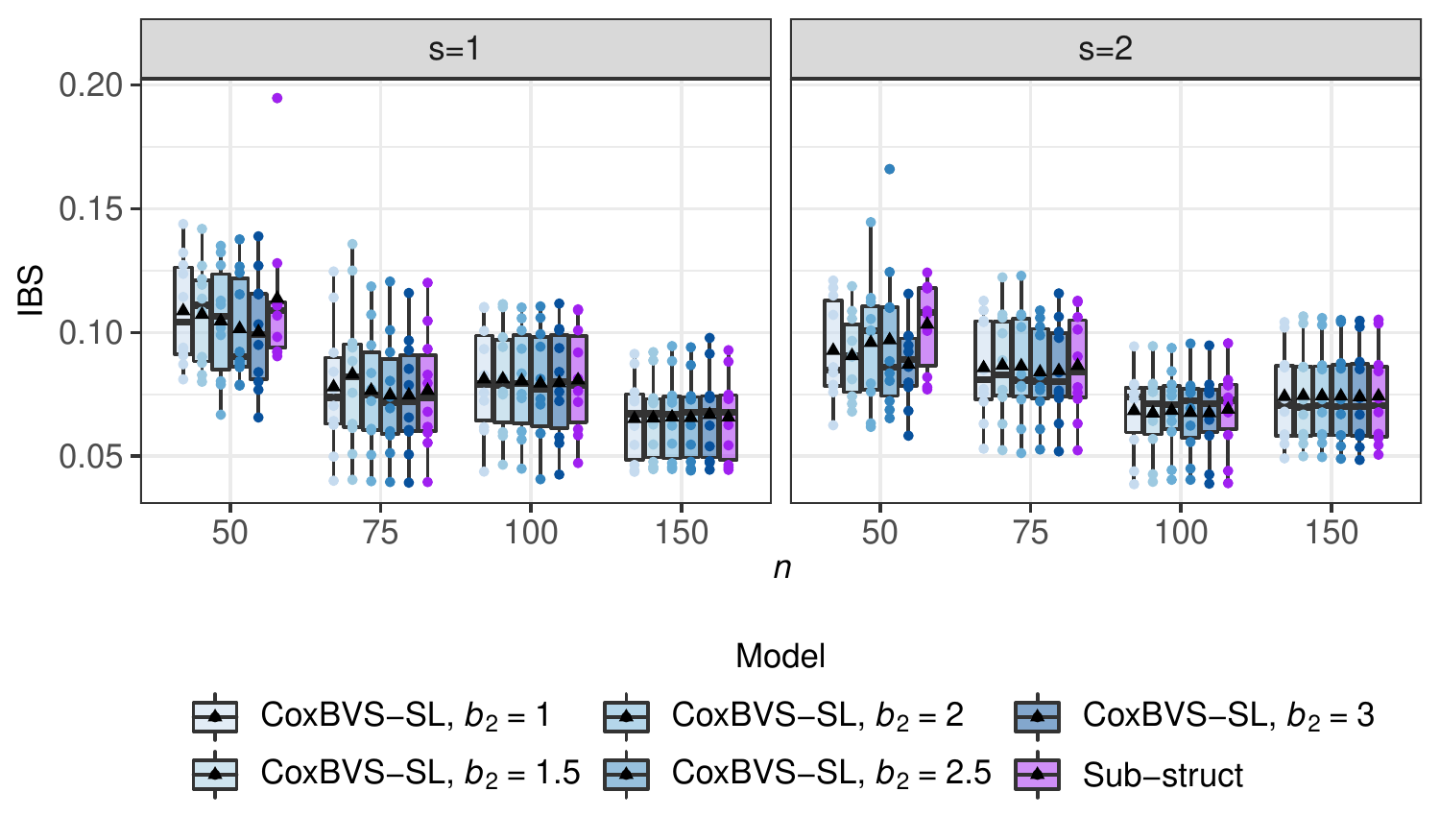} 

}

\end{knitrout}
\caption{Integrated Brier Scores (IBS) across all ten test sets for subroup~1 (left) and~2 (right) (based on the Bayesian Model Averaging). The black triangle within each boxplot represents the mean value.} 
\label{fig:BayesSim2IBSBMA}
\end{figure}

\begin{figure}[!htb] 
\begin{knitrout}
\definecolor{shadecolor}{rgb}{0.969, 0.969, 0.969}\color{fgcolor}

{\centering \includegraphics[width=\linewidth]{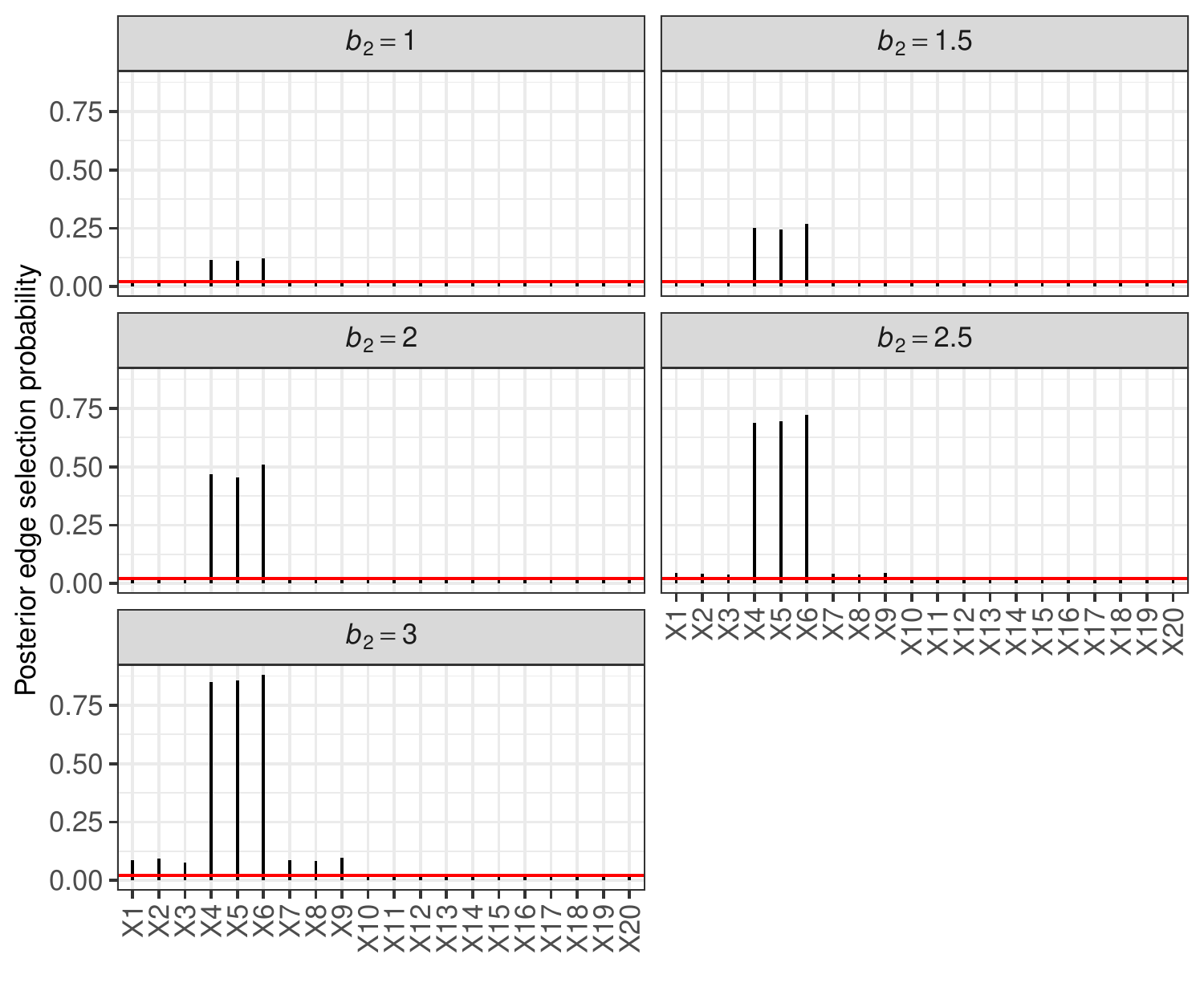} 

}

\end{knitrout}
\caption{Marginal posterior edge selection probabilities (averaged across all trai\-ning sets) of the first 20 genes in $\boldsymbol{G}_{12}$ and $n=p=100$. The red line indicates the prior mean ($\pi=2/(p-1) \approx 0.02$ for $p=100$).} 
\label{fig:BayesSim2G12}
\end{figure}


\begin{figure}[!htb]
\begin{knitrout}
\definecolor{shadecolor}{rgb}{0.969, 0.969, 0.969}\color{fgcolor}

{\centering \includegraphics[width=\linewidth]{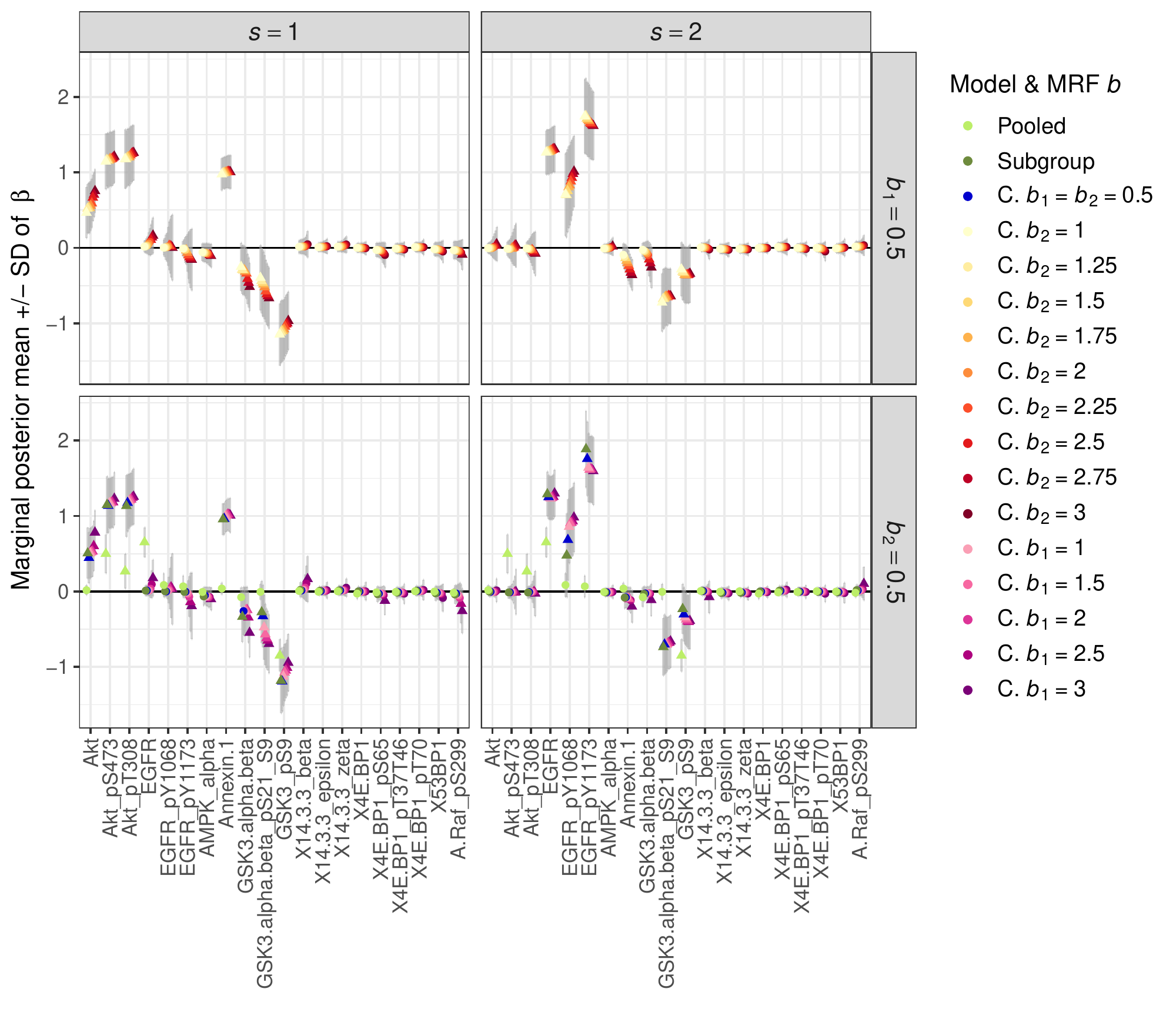} 

}

\end{knitrout}
\caption{Marginal posterior means (independent of $\gamma = 1$) and standard deviations (SD)  of the regression coefficients of all 20 proteins in both subgroups (averaged across all training sets). The different colors represent the models or parameter values of $b_1$ and $b_2$ in CoxBVS-SL (abbreviated by "C."). The plot symbol indicates whether a protein is selected (triangle) or not (circular point).} 
\label{fig:GBmargBetas}
\end{figure}

\begin{figure}[!htb]
\begin{knitrout}
\definecolor{shadecolor}{rgb}{0.969, 0.969, 0.969}\color{fgcolor}

{\centering \includegraphics[width=\linewidth]{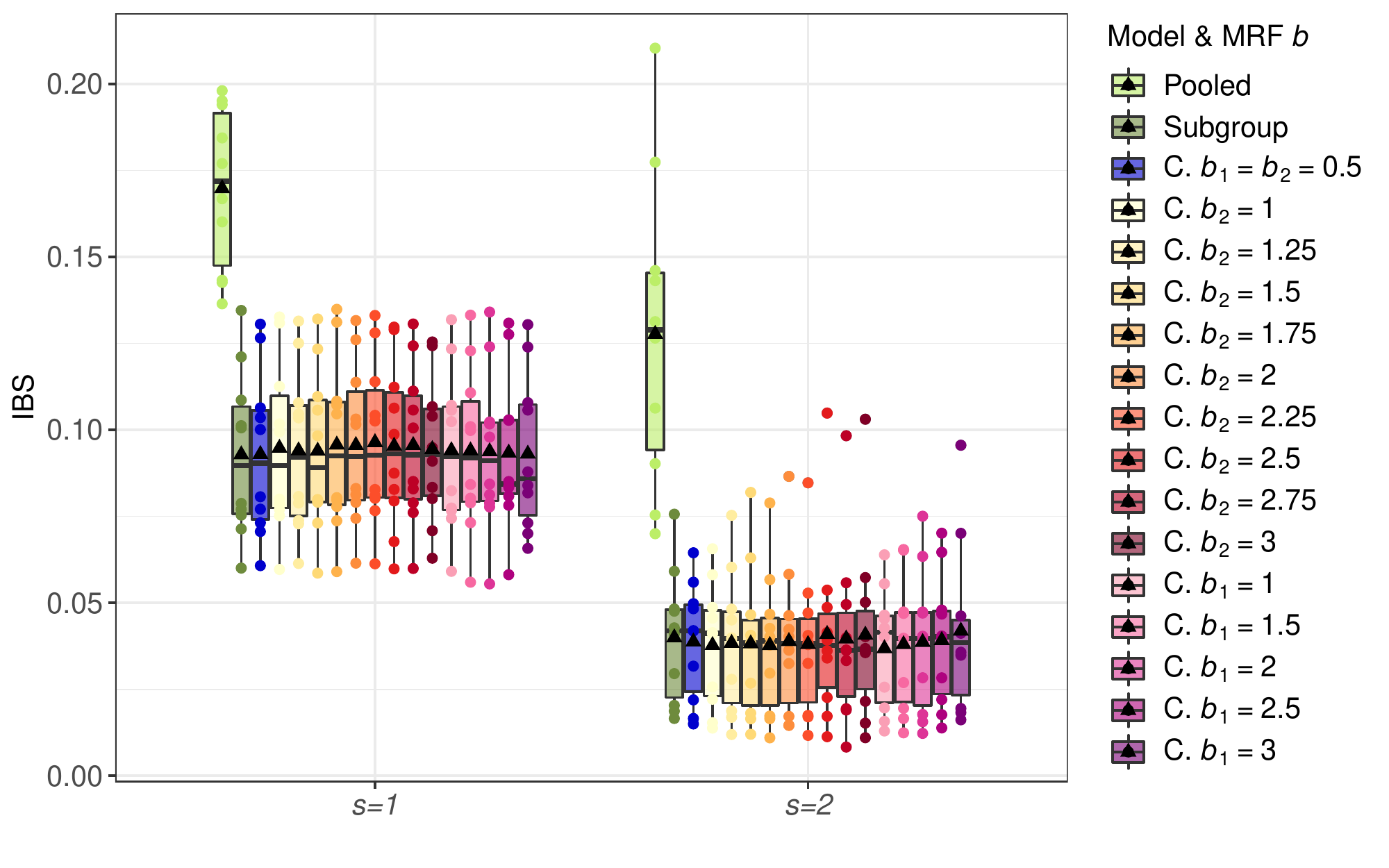} 

}

\end{knitrout}
\caption{Integrated Brier Scores (IBS) across all ten test sets for both subroups (based on the Bayesian Model Averaging). CoxBVS-SL is abbreviated by "C.". The black triangle within each boxplot represents the mean value.} 
\label{fig:GBibsBMA}
\end{figure}

\begin{figure}[!htb] 
\begin{knitrout}
\definecolor{shadecolor}{rgb}{0.969, 0.969, 0.969}\color{fgcolor}

{\centering \includegraphics[width=\linewidth]{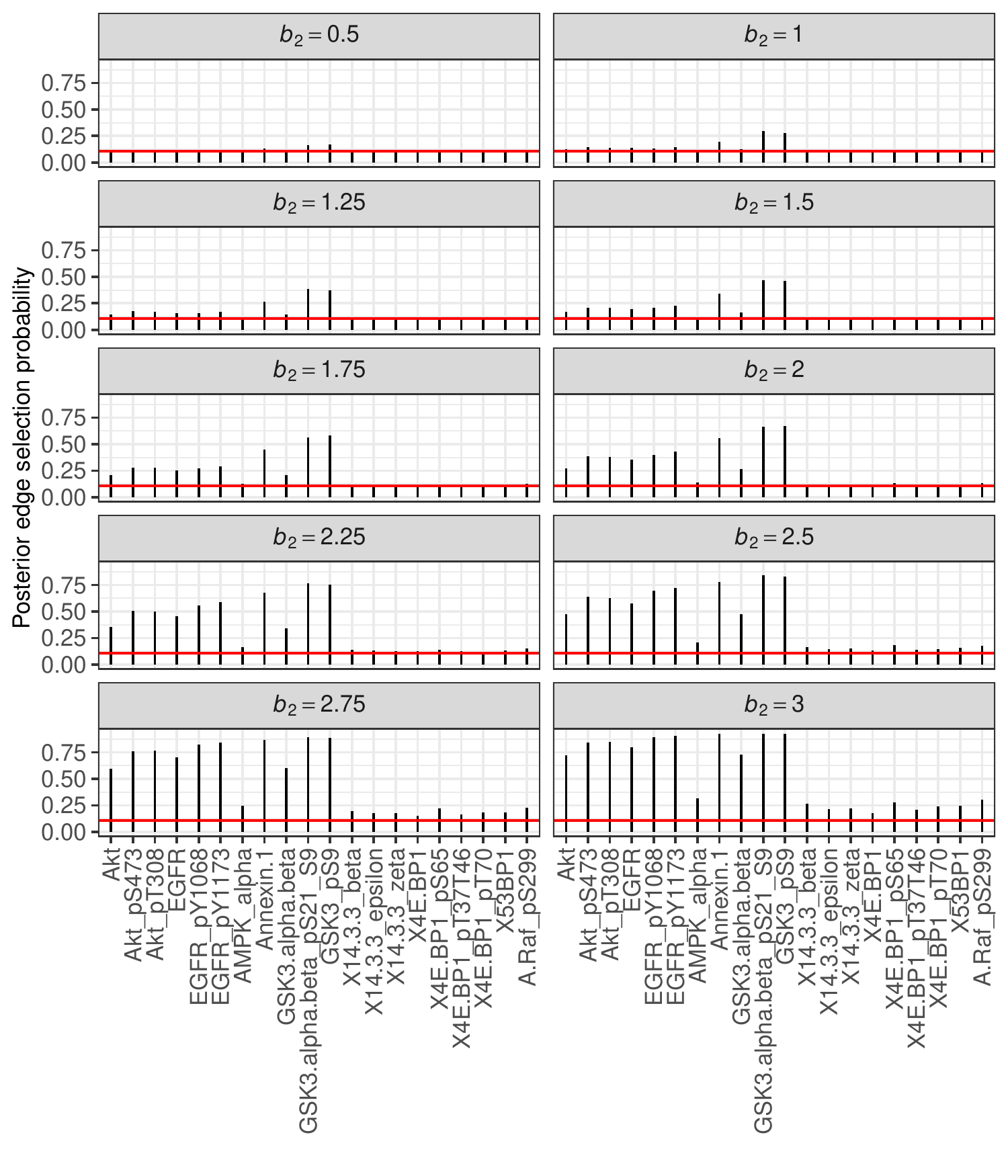} 

}

\end{knitrout}
\caption{Mean marginal posterior edge selection probabilities for $\boldsymbol{G}_{12}$ (averaged across all training sets) in the CoxBVS-SL model with $b_1=0.5$. The red line indicates the prior mean ($\pi=2/(p-1) \approx 0.11$ for $p=20$).} 
  \label{fig:GBgraphG12}
\end{figure}

\end{document}